\documentclass[11pt]{article}

\usepackage[margin=1in]{geometry}
\usepackage{amsmath,amsfonts,amssymb,amsthm}
\usepackage{mathrsfs,mathtools}
\usepackage{bbm,bm}
\usepackage[shortlabels]{enumitem}
\usepackage{microtype}
\usepackage{booktabs,array}
\usepackage{algorithm}
\usepackage{algpseudocode}
\usepackage[dvipsnames]{xcolor}
\usepackage{hyperref}
\usepackage{marginnote}

\usepackage{authblk}

\hypersetup{
  pdftitle={Half-Optimal Induced Matchings in Random Graphs},
  pdfkeywords={induced matchings, random graphs, online algorithms, look-ahead, low-degree polynomials},
  colorlinks=true,
  linkcolor=blue!55!black,
  citecolor=blue!55!black,
  urlcolor=blue!65!black
}

\newtheorem{theorem}{Theorem}[section]
\newtheorem{proposition}[theorem]{Proposition}
\newtheorem{lemma}[theorem]{Lemma}

\theoremstyle{definition}
\newtheorem{definition}[theorem]{Definition}

\theoremstyle{remark}
\newtheorem{remark}[theorem]{Remark}

\newcommand{\Prb}{\mathbb P}
\newcommand{\RR}{\mathbb R}
\newcommand{\NN}{\mathbb N}
\newcommand{\E}{\mathbb E}

\newcommand{\cA}{\mathcal A}
\newcommand{\cB}{\mathcal B}
\newcommand{\cF}{\mathcal F}
\newcommand{\Bin}{\mathrm{Binomial}}
\newcommand{\Mout}{M_{\mathrm{out}}}

\newcommand{\bbone}{\mathbbm 1}
\newcommand{\abs}[1]{\left|#1\right|}
\newcommand{\norm}[1]{\left\lVert#1\right\rVert}

\newcommand{\inparen}[1]{\left(#1\right)}
\newcommand{\inbraces}[1]{\left\{#1\right\}}
\newcommand{\insquare}[1]{\left[#1\right]}

\newcommand{\ER}{Erd\H{o}s--R\'enyi }
\newcommand{\eps}{\varepsilon}

\newcommand{\ab}{\mathbf a}
\newcommand{\bbb}{\mathbf b}
\newcommand{\tb}{\mathbf t}
\newcommand{\Sbb}{\mathbf S}
\counterwithin{algorithm}{section}

\makeatletter
\newenvironment{breakablealgorithm}
  {
   \begin{center}
     \refstepcounter{algorithm}
     \hrule height.8pt depth0pt \kern2pt
     \renewcommand{\caption}[2][\relax]{
       {\raggedright\textbf{\ALG@name~\thealgorithm} ##2\par}%
       \ifx\relax##1\relax 
         \addcontentsline{loa}{algorithm}{\protect\numberline{\thealgorithm}##2}%
       \else 
         \addcontentsline{loa}{algorithm}{\protect\numberline{\thealgorithm}##1}%
       \fi
       \kern2pt\hrule\kern2pt
     }
  }{
     \kern2pt\hrule\relax
   \end{center}
  }
\makeatother

\title{Half-Optimal Induced Matchings in Random Graphs:\\
Online and Low-Degree Algorithms}

\author[1]{Abhishek Dhawan\thanks{Email: \textit{adhawan2@illinois.edu}. Partially supported by the NSF RTG grant DMS-1937241.}}
\author[2]{Eren C. K{\i}z{\i}lda\u{g}\thanks{Email: \textit{kizildag@illinois.edu}.}}
\author[1]{Neeladri Maitra\thanks{Email: \textit{nmaitra@illinois.edu}.}}
\author[3]{Cheng Mao\thanks{Email: \textit{cheng.mao@math.gatech.edu}. Partially supported by NSF CAREER Award 2338062.}}
\author[3]{Timothy~L.~H.~Wee\thanks{Email: \textit{timothy.wee@gatech.edu}.}}

\makeatletter
\@namedef{@sep4}{,\authorcr}
\@namedef{@sep5}{\Authand}
\makeatother

\affil[1]{Department of Mathematics, University of Illinois Urbana-Champaign}
\affil[2]{Department of Statistics, University of Illinois Urbana-Champaign}
\affil[3]{School of Mathematics, Georgia Institute of Technology}

\date{}

\begin{document}

\maketitle

\begin{abstract}
For many optimization problems on random graphs, most prominently the largest independent set problem, natural algorithms reach only half the optimum, and such \emph{factor-two gaps} are known to be sharp within broad algorithmic classes. We prove such a gap for finding large \emph{induced matchings} in \ER random graphs. In the worst case, requiring a matching to be induced turns a polynomial-time solvable problem into an NP-hard one; our results reveal an average-case counterpart.

In $G(n,p)$ with $C/n\le p\le 1-\kappa$, the largest induced matching has size $(1+o(1))L$ with high probability, where $L=\log_{1/(1-p)}(np)$. A simple online greedy algorithm finds an induced matching of size $(1/2-\eps)L$, while no online algorithm, even with adaptive inspection order and unbounded computation, finds one of size $(1/2+\eps)L$ except with probability $\exp(-\Omega(L\log(np)))$. In the sparse regime $p=d/n$, low-degree polynomial algorithms exhibit the same sharp threshold.

Our guarantees rest on a \emph{maximality principle}: with high probability, every maximal induced matching is half-optimal. Greedy outputs a maximal induced matching and our low-degree algorithm a near-maximal one. We bypass the delicate coupling arguments classically used to analyze greedy since Karp. For low-degree hardness, we adapt Wein's overlap gap argument, with forbidden structures that track overlaps between endpoint sets rather than edge sets, since a new edge may join two previously used endpoints and thus encounter no fresh randomness. For online hardness, we give a new and significantly simpler argument that avoids the correlated instances of prior OGP-based proofs: at a suitable stopping time, all pairs involving uninspected vertices remain unrevealed, so a direct first-moment bound suffices.
\end{abstract}

\pagenumbering{gobble}
\newpage
\tableofcontents
\newpage
\pagenumbering{arabic}
\section{Introduction}
\label{sec:introduction}
A central feature of many optimization problems on random graphs is an \emph{optimization-to-computation gap}: known polynomial-time algorithms find solutions substantially smaller than those guaranteed to exist. Perhaps the earliest and most prominent example is the largest independent set problem in the \ER random graph $G(n,1/2)$. Non-constructive moment arguments show that the largest independent set has size $(2+o(1))\log_2 n$, while a simple greedy algorithm finds one of size $(1+o(1))\log_2 n$, both with high probability (whp)~\cite{grimmett1975colouring}. In his seminal 1976 work, Karp challenged the community to find a polynomial-time algorithm that produces an independent set of size $(1+\eps)\log_2 n$ whp for some fixed $\eps>0$~\cite{karp1976probabilistic}. Karp's question remains one of the most prominent open problems in the algorithmic theory of random graphs, and was highlighted by Frieze in his 2014 ICM talk~\cite{F2014}: no polynomial-time algorithm is known to improve on the leading constant achieved by greedy. Understanding such gaps has fueled a rich line of work, revealing analogous \emph{factor-two gaps} across a range of random optimization problems; see, e.g.,~\cite{achlioptas2008algorithmic,coja2015independent,Wein2021,GamarnikKizildagWarnke2025}.

This line of work has recently been extended to problems with global constraints. Perkins and Wang~\cite{perkins2024hardness} established a factor-two gap for local and low-degree algorithms that find \emph{balanced} independent sets in sparse random bipartite graphs, even though the unconstrained maximum independent set problem on bipartite graphs is solvable in polynomial time. Their work is motivated by statistical-mechanics heuristics suggesting that global constraints can induce \emph{glassy} behavior and computational barriers in otherwise tractable problems~\cite{mezard1987mean,mezard2010glassy}. These developments raise a fundamental question: how do algorithmic gaps depend on the structure required of a solution? Induced matchings provide a natural setting in which to address this question beyond independent sets.

Given a graph, an \emph{induced matching} (also known as a \emph{strong matching}) is a collection of vertex-disjoint edges whose endpoints span no other edges. In this paper, we study the algorithmic problem of finding large induced matchings in \ER random graphs. Throughout, the size of a matching is its number of edges. Let $G\sim G(n,p)$ be an \ER random graph with vertex set $[n]\coloneqq \{1,\dots,n\}$, in which each unordered pair of vertices forms an edge independently with probability $p$. Set
\begin{equation}\label{eq: def of b and L}
    b\coloneqq b(p)=\frac{1}{1-p}, \qquad L\coloneqq \log_b(np) = \frac{\log(np)}{-\log(1-p)}.
\end{equation}
Our main results concern the regime $C/n\le p\le 1-\kappa$, where $\kappa\in(0,1)$ is fixed and $C>0$ is sufficiently large. In particular, they cover the \emph{sparse} and \emph{dense} regimes as well as the \emph{intermediate} range $1/n\ll p\ll 1$.

For a broad range of densities $p$, the size of the largest induced matching is known to first order. Cooley, Dragani{\'c}, Kang, and Sudakov proved that for every $\xi>0$ there is a $C(\xi)>0$ such that, whp, this size lies between $(1-\xi)L$ and $(1+\xi)L$, uniformly for all $C(\xi)/n\le p\le 0.99$~\cite[Theorem~1.1]{CooleyDraganicKangSudakov2021}. In particular, the optimum is $(1+o(1))L$ whp when $np\to\infty$ in this range. For constant $p\in(0,1)$, sharper estimates including lower-order terms were established earlier~\cite{el1994maximum,clark2001strong}.

Induced matchings are the simplest induced packings with nontrivial components, and thus a basic instance of the problem of finding large induced subgraphs with prescribed structure. Their algorithmic study, however, presents distinct challenges: an algorithm must select both the endpoints and their pairing, while ensuring that distinct pairs remain mutually isolated. The standard reduction to independent sets does not resolve these difficulties. Although induced matchings correspond to independent sets in the square of the line graph (see below), this transformation produces a random graph with strong dependencies. Sharp algorithmic results for independent sets therefore do not directly determine the algorithmic limits for induced matchings, which leads to the question motivating this work:
\begin{center}
\emph{How large an induced matching can natural classes of algorithms find?}
\end{center}

We investigate this question in two frameworks: online algorithms and low-degree polynomials. An \emph{online} algorithm inspects the vertices one at a time and, upon inspecting a vertex, learns its adjacencies to previously inspected vertices. It may then add an edge joining the current vertex to an earlier one, provided that the accepted edges still form an induced matching. Each vertex may be inspected at most once, accepted edges are irrevocable, and the inspection order may be adaptive. In particular, an inspected vertex may remain available until the algorithm pairs it with a later vertex. Thus, the restriction concerns access to the graph and the timing of decisions; computation and memory are otherwise unrestricted. This model also shows why the reduction to independent sets does not preserve online structure: inspecting a single vertex of $G$ reveals an entire collection of candidate edges, corresponding to vertices of the square of the line graph, among which the algorithm may choose after seeing all newly revealed adjacencies. Online algorithms have been the focus of a growing body of recent work in the algorithmic theory of random graphs~\cite{GamarnikKizildagWarnke2025,GamarnikRaczSchoenbach2026,dhawan2025sharp,dhawan2026algorithmic}.

In contrast to the online setting, a \emph{low-degree polynomial} (LDP) algorithm has access to the entire adjacency matrix. It assigns a real-valued score to each potential edge, where each score is a polynomial of bounded total degree in the adjacency indicators. A prescribed rounding procedure then converts these scores into an induced matching. Following the low-degree optimization framework of~\cite{Wein2021}, we impose a normalization on the second moment of the score vector and a bound on the rounding error. Randomization is allowed through polynomial coefficients that depend on a seed independent of the graph. %
Here, the algorithmic restriction concerns the algebraic complexity of the scores, as measured by their total degree. The online and low-degree frameworks therefore capture fundamentally different algorithmic constraints.

Our results show that, in both frameworks, algorithms find half-optimal induced matchings and can do no better. In particular, a simple online greedy algorithm finds an induced matching of size at least $(1/2-\eps)L$ whp throughout the range $C/n\le p\le 1-\kappa$, where $C>0$ is sufficiently large. Moreover, this guarantee is optimal within the online model, even with adaptive inspection order and unrestricted computation. For sparse random graphs $G(n,d/n)$, a constant-degree polynomial finds an induced matching of size approximately $n\log d/(2d)$ in the double limit $n\to\infty$ followed by $d\to\infty$. We complement this construction with a matching low-degree hardness bound that gives an explicit trade-off between polynomial degree and success probability. Taken together, our results reveal factor-two computational gaps for induced matchings, extending the \emph{half-optimality} phenomenon to a new optimization problem and providing further evidence of its prevalence across algorithmic frameworks.

Our arguments introduce new tools for this problem, notably (i) a \emph{maximality} principle for establishing algorithmic guarantees, and (ii) a direct proof of sharp lower bounds for online algorithms that avoids the construction of correlated instances used in prior work~\cite{GamarnikKizildagWarnke2025,GamarnikRaczSchoenbach2026,dhawan2025sharp,dhawan2026algorithmic}.

\paragraph{Comparison with independent sets} 

We now give a heuristic explanation for the hardness of finding near-maximum induced matchings, based on the typical structure of independent sets in locally sparse graphs.

An induced matching of size $m$ is equivalently a pair of disjoint independent sets $I$ and $J$, each of size $m$, such that $G[I\cup J]$ is a perfect matching.
Since the largest induced matching in $G(n,p)$ has size $(1+o(1))L$ whp, a near-maximum induced matching consists of two independent sets of size about $L$ whose union, of size about $2L$, is comparable to the independence number $(2+o(1))L$.
It is therefore natural to ask whether a \emph{typical} independent set can serve as one of the two sides.

The structure of typical independent sets in sparse graphs is well understood.
Davies, Jenssen, Perkins, and Roberts~\cite{davies2018average} showed that, in an $n$-vertex triangle-free graph of maximum degree $\Delta$, the average size of an independent set is at least $(1-o(1))\,n\log \Delta / \Delta$.
Their argument, which builds on an approach introduced by Shearer~\cite{shearer1995independence} and further developed by Alon~\cite{alon1996independence}, in fact yields more: a typical independent set $I$ has size at least $(1-o(1))\,n\log \Delta / \Delta$, and most vertices outside $I$ have at least $(1-o(1))\log \Delta$ neighbors in $I$.
These conclusions transfer to $G(n,d/n)$ when $\log n \ll d \ll n^{1/3}$. In this regime, all degrees are $(1+o(1))d$ whp, and deleting the vertices that lie in triangles yields a triangle-free induced subgraph on $(1-o(1))n$ vertices.
Thus a typical independent set in $G(n,d/n)$ has size about $L=(1+o(1))\,n\log d/d$, precisely the size of one side of a near-maximum induced matching, and a typical vertex outside it has about $\log d$ neighbors in it.

Now let $I$ be a typical independent set of size $L$ in $G(n,d/n)$.
For $I$ to be one side of an induced matching of size $L$, there must be a set $J$ of $L$ vertices such that
\begin{enumerate}[label=(\roman*)]
    \item\label{cond: j induced matching} every vertex of $J$ has exactly one neighbor in $I$,
    \item distinct vertices of $J$ have distinct neighbors in $I$, and
    \item\label{cond: j independent} $J$ is an independent set.
\end{enumerate}
We show heuristically that conditions~\ref{cond: j induced matching} and~\ref{cond: j independent} are already incompatible.
Treating the number of neighbors in $I$ of a vertex $v\notin I$ as a $\mathrm{Poisson}(\lambda)$ random variable with $\lambda = d|I|/n \approx \log d$, the set $U$ of vertices with exactly one neighbor in $I$ has size about
\[
    n\cdot \log d\cdot e^{-\log d} = \frac{n\log d}{d} \approx L.
\]
Condition~\ref{cond: j induced matching} therefore forces $J$ to consist of nearly all of $U$.
On the other hand, $U$ is defined only through the edges between $U$ and $I$, so the edges within $U$ are, heuristically, unexposed. Thus $G[U]$ behaves like a random graph on $|U|$ vertices with edge probability $d/n$, and hence with average degree about $|U|d/n \approx \log d$.
The independence number of such a graph is $(2+o(1))\,|U|\log\log d/\log d = o(L)$, far below the size $L$ required by condition~\ref{cond: j independent}.

Consequently, a typical independent set of size $L$ is very unlikely to be one side of a near-maximum induced matching. This suggests that near-maximum induced matchings are built from atypical independent sets, which natural algorithms are not expected to find.

\paragraph{Hard problems hidden inside easy problems}
Our choice of the induced matching problem is motivated by the phenomenon of \emph{hidden hardness}: a problem may be efficiently solvable, while finding a solution subject to an additional constraint is computationally hard. Maximum matchings can be found in polynomial time~\cite{Edmonds1965Paths}, but imposing the \emph{inducedness} condition makes the problem NP-hard~\cite{cameron1989induced}. Our results give an average-case counterpart: in random graphs, natural classes of algorithms fail to find induced matchings of more than half the maximum size. A similar phenomenon occurs for large \emph{balanced} independent sets in random bipartite graphs and multipartite hypergraphs (see~\cite{perkins2024hardness, dhawan2026low} for the sparse case and~\cite{dhawan2025sharp, dhawan2026algorithmic} for the dense case). There, finding a large independent set is easy, since one may simply take one of the parts, but imposing balancedness induces algorithmic phase transitions for natural classes of algorithms. Hidden hardness has also been observed, in both worst-case and average-case forms, in several other settings:
\begin{itemize}
    \item The partition function of the ferromagnetic Ising model with a consistent external field can be approximated efficiently at all temperatures and on all graphs by the randomized algorithm of Jerrum and Sinclair~\cite{jerrum1993polynomial}. In contrast, Carlson, Davies, Kolla, and Perkins~\cite{carlson2022computational} showed that fixing the magnetization makes approximating the resulting partition function computationally hard in certain regimes, giving rise to computational thresholds.
    \item For constraint satisfaction problems (CSPs), Bulatov and Marx~\cite{bulatov2010complexity} classified exactly which constraint languages remain polynomial-time solvable under global cardinality constraints; in particular, such constraints can turn tractable CSPs into NP-complete ones.
    \item A densest subgraph of a graph can be found in polynomial time~\cite{goldberg1984finding, lawler2001combinatorial}, whereas finding a densest subgraph under a constraint on its number of vertices is NP-hard~\cite{khuller2009finding}.
\end{itemize}
Our results on induced matchings add to this growing list of easy problems with hard constrained variants. Given how varied these settings are, it is natural to ask whether hidden hardness is \emph{universal} in some sense: when an easy problem becomes hard under an additional constraint, can one say something general about the source or nature of that hardness?

\subsection{Overview of contributions}
We now informally summarize our main contributions. Below, whp refers to $n\to\infty$, with probabilities taken over the graph and any internal randomness of the algorithm.

\paragraph{Online algorithms} Our first results concern online algorithms, which inspect the graph sequentially and make irrevocable decisions about which edges to accept, over a broad range of densities; see Section~\ref{sec:models-results} for the formal model. We establish a guarantee for a simple greedy algorithm and a matching impossibility bound for all online algorithms.
\begin{theorem}[Informal, see Theorems~\ref{thm:greedy-main} and~\ref{thm:broad-main}]
For every $\eps\in(0,1/2)$ and $\kappa\in(0,1)$, there exists $C=C(\eps,\kappa)>0$ such that the following holds uniformly for $C/n\le p\le 1-\kappa$. A simple online greedy algorithm finds an induced matching of size at least $(1/2-\eps)L$ whp. Conversely, every online algorithm in our model finds an induced matching of size at least $(1/2+\eps)L$ with probability $o(1)$.
\end{theorem}
Importantly, this result imposes no restriction on running time and holds for randomized algorithms and arbitrary inspection orders. The $o(1)$ term in our impossibility result is $\exp(-\Omega(L\log (np)))$, which is optimal up to the constant in the exponent: the online algorithm that inspects a fixed set of $2\lceil(1/2+\eps)L\rceil$ vertices and accepts every edge it encounters succeeds whenever these vertices induce a perfect matching, which happens with probability $\exp(-O(L\log (np)))$. The greedy algorithm maintains an induced matching $M$ and an independent buffer $B$ of inspected, unmatched vertices with no neighbors among the endpoints of $M$. Upon inspecting a vertex $v$, it discards $v$ if $v$ has a neighbor among these endpoints. Otherwise, if $v$ has no neighbor in $B$, it adds $v$ to $B$; if it does have a neighbor in $B$, it picks any such neighbor $u$, accepts the edge $uv$ and includes it in the current induced matching, and removes from $B$ all neighbors of $v$, including $u$.

\paragraph{Low-degree bounds in sparse graphs}
We next turn to low-degree polynomial (LDP) algorithms in sparse random graphs $G(n,d/n)$. Set $\Phi\coloneqq n\log d/d$.
\begin{theorem}[Informal, see Theorems~\ref{thm:sparse-positive-low-degree} and~\ref{thm:sparse_low_degree_hardness}]
For every $\eps\in(0,1/2)$ and sufficiently large $d$, a constant-degree %
polynomial produces an induced matching of size at least $(1/2-\eps)\Phi$ whp. Conversely, for $1\le D\le cn/\log n$, no randomized degree-$D$ polynomial satisfying our normalization and rounding conditions produces an induced matching of size at least $(1/2+\eps)\Phi$ with probability at least $1-\exp(-CD\log n)$, where $c,C>0$ depend on $\eps$, $d$, and the fixed normalization bound.
\end{theorem}
Our formal result gives an explicit trade-off between the degree and the success probability that is ruled out, including the dependence on the normalization and rounding error. Our construction works with the square of the line graph of $G$, whose independent sets are precisely the induced matchings of $G$. We show that, after truncating large neighborhoods, a local algorithm on this graph can be approximately simulated by constant-degree polynomials; see Section~\ref{sec:positive-low-degree}.
\begin{remark}
    Our low-degree results concern sparse random graphs, while our online results cover a broader range of densities. A similar distinction arises for independent sets. Although low-degree hardness bounds are known for dense graphs~\cite[Section~1.3]{Wein2021}, whether low-degree polynomials can match the online greedy guarantee remains open. In fact, for $G(n,1/2)$, polynomials of degree $o(\log^2 n)$ are conjectured to fail to find independent sets of size $0.9\log_2 n$, even though online greedy finds ones of size $(1-o(1))\log_2 n$; see~\cite{AIM2024}. In the intermediate regime $1/n\ll p\ll 1$, we are not aware of any low-degree results, either achievability or hardness.
\end{remark}

\subsection{Proof techniques}
\label{subsec:proof-techniques}
In this section, we highlight the key technical ideas underlying our results.

\paragraph{Online achievability}
A central concept in our work is \emph{maximality}: an induced matching is maximal if no further edge can be added while preserving the induced matching property. Since the greedy algorithm returns a maximal induced matching by its very construction, it suffices to show that every maximal induced matching is large.

To this end, fix a candidate matching with endpoint set $S$, and let $R_S$ be the set of vertices outside $S$ with no neighbor in $S$. Conditional on $S$ spanning an induced matching, maximality is equivalent to $G[R_S]$ being edgeless. For a candidate of size $k\le (1/2-\eps)L$, the random variable $|R_S|$ is binomial with mean $(n-2k)(1-p)^{2k} = \Omega\bigl((np)^{2\eps}/p\bigr)$. Concentration of $|R_S|$, together with the requirement that $G[R_S]$ be edgeless, makes maximality sufficiently unlikely. A first-moment bound over all candidates then rules out such small maximal induced matchings whp, which establishes the desired guarantee for greedy.

This approach avoids the involved coupling arguments used in prior analyses of greedy algorithms (see, e.g.,~\cite{karp1976probabilistic, GamarnikKizildagWarnke2025, dhawan2025sharp, dhawan2026algorithmic}): rather than tracking the dynamics of the algorithm, we show that \emph{every} maximal induced matching is large. We expect this to be useful for analyzing greedy algorithms in other settings, since such algorithms by construction output maximal structures.

\paragraph{Online lower bound}
Unlike prior work~\cite{GamarnikKizildagWarnke2025,GamarnikRaczSchoenbach2026,dhawan2025sharp,dhawan2026algorithmic}, our proof of the lower bound for online algorithms does \emph{not} construct temporally evolving families of correlated instances. Instead, we work directly with a single exposure of the graph. Let $p_{\rm suc}$ denote the probability that an online algorithm finds an induced matching of size $m^*\coloneqq (1/2+\eps)L$, and set $r\coloneqq (1/2+\eps/2)L$ and $s\coloneqq m^*-r=\eps L/2$. Let $\mathcal{A}_t$ be the matching accepted after $t$ inspections, and define $\tau_r\coloneqq \min\{t\ge 1:|\mathcal{A}_t|\ge r\}$, the first time the matching reaches size $r$. Since the algorithm adds at most one edge per inspection, $|\mathcal{A}_{\tau_r}|=r$ whenever $\tau_r<\infty$.

Condition on the full history $\mathcal{F}_{\tau_r}$ on the event $\tau_r=t<\infty$, and let $S$ be the set of $2r$ endpoints of $\mathcal{A}_{\tau_r}$. On the success event, the algorithm must accept at least $s$ further edges. Each of these edges has a newly inspected endpoint, and $s$ of these endpoints form a set $X$ of vertices that are still uninspected at time $\tau_r$. Since the final matching is induced and decisions are irrevocable, $X$ can be chosen to be an independent set with no edges to $S$. This forbids a total of $\binom{s}{2}+2rs$ pairs. For each fixed candidate $X$, all of these pairs are unrevealed at time $\tau_r$, so their edge indicators remain conditionally independent ${\rm Bernoulli}(p)$ variables. A first-moment bound over all choices of $X$ therefore gives
\[
\mathbb{P}\bigl[
\text{at least $s$ further edges are accepted}
\mid \mathcal{F}_{\tau_r}
\bigr]
\le \binom{n}{s}(1-p)^{2rs+\binom{s}{2}}.
\]
For sufficiently large $C=C(\eps,\kappa)$, our choice of parameters makes the right-hand side at most $\exp\bigl(-\eps^2\log^2(np)/(2\log b)\bigr)$. Since this estimate holds uniformly over the conditioning, the same upper bound applies to $p_{\rm suc}$.

\paragraph{Low-degree achievability}
Turning to $G(n,d/n)$, we now outline the proof of our low-degree achievability result. Given a graph $G$, let $L(G)$ be its line graph: its vertices are the edges of $G$, and two distinct vertices are adjacent if the corresponding edges share an endpoint in $G$. We consider the square of the line graph, $H=L(G)^2$, in which two distinct vertices are adjacent if their distance in $L(G)$ is at most two. Equivalently, two edges of $G$ are adjacent in $H$ if they share an endpoint or if some edge of $G$ joins an endpoint of one to an endpoint of the other; we call such edges \emph{conflicting}. Thus, induced matchings in $G$ correspond precisely to independent sets in $H$.

We use this observation to build a local algorithm and simulate it by a constant-degree polynomial. Specifically, we fix a sufficiently large integer $Q\coloneqq Q(d,\eps)$ and assign each potential edge an independent uniform color in $\{1,\dots,Q\}$, independently of $G$. These colors induce a vertex coloring of $H$. We delete every vertex of $H$ that has a neighbor of the same color, and denote the set of deleted vertices by $C$. Each color class is then an independent set in $H\setminus C$. We run greedy in $Q$ \emph{color rounds}: in round $j$, we accept every remaining vertex (in the square of the line graph) of color $j$ that has no neighbor accepted in an earlier round. Since vertices accepted in the same round share a color, they are not adjacent; hence the output $I$ is an independent set in $H$, and thus an induced matching in $G$.

Moreover, $I$ is maximal in $H\setminus C$. Extend $I$ to a maximal independent set $J$ of $H$. Every vertex outside $I\cup C$ already has a neighbor in $I$, so every added vertex belongs to $C$. Consequently, $|I|\ge |J|-|C|$. Choosing $Q$ sufficiently large makes $|C|$ small whp, so our lower bound on the size of every maximal induced matching yields the desired half-optimal guarantee for $I$.

This algorithm is local: determining whether a vertex is accepted requires recursively examining decisions only at neighbors of smaller colors. Such dependency chains have length at most $Q-1$, so each decision depends only on a neighborhood of bounded radius. We then fix a sufficiently large constant $K=K(d,\eps)$ and construct polynomials of degree at most $K$ that agree with the local rule whenever the relevant neighborhood contains at most $K$ edges. Tail bounds on neighborhood sizes control the rounding loss, while moment estimates give the required second-moment normalization. Finally, an averaging argument fixes the colors, yielding a deterministic polynomial whose degree is independent of $n$. Full details appear in Section~\ref{sec:positive-low-degree}.

\paragraph{Low-degree lower bound}
Our low-degree lower bound is based on a variant of the asymmetric overlap gap property (OGP), which rules out certain tuples of solutions, called \emph{forbidden tuples}. We show that a sufficiently successful low-degree algorithm would produce such a tuple, a contradiction. An asymmetric OGP argument also underlies the optimal low-degree hardness result for the maximum independent set problem in sparse random graphs~\cite{Wein2021}. A fundamental difference is that our forbidden tuples track overlaps between endpoint sets rather than between the matching edge sets themselves.

Specifically, recall $\Phi=n\log d/d$ and set $k_*\coloneqq (1/2+\eps)\Phi$. Suppose a low-degree algorithm finds an induced matching of size at least $k_*$ with sufficiently high probability. Consider a sequence of correlated graphs $G^{(0)},G^{(1)},\dots,G^{(T)}$ obtained by resampling one edge indicator at a time. A forbidden tuple consists of $K$ matchings $S_1,\dots,S_K$, where each $S_j$ is an induced matching in some $G^{(t_j)}$, such that
\begin{itemize}
\item $|S_j|\ge k_*$ for all $j=1,\dots,K$, and
\item $\frac{\eps}{4}\Phi \le \bigl|V(S_j)\setminus\bigcup_{\ell<j}V(S_\ell)\bigr| \le \frac{\eps}{2}\Phi$ for all $j=2,\dots,K$, where $V(S_j)$ denotes the endpoint set of $S_j$.
\end{itemize}
Thus, each successive matching is large but introduces only a prescribed, small number of new endpoints. For a suitable $K=O(\eps^{-2})$ and all sufficiently large fixed $d$, a first-moment argument shows that, except with probability $\exp(-\Omega(n))$, no such tuple exists along the path; see Proposition~\ref{prop:sparse_endpoint_forbidden_structure}. Tracking new endpoints, rather than new edges, is crucial. The earlier induced-matching events constrain only adjacency indicators within their endpoint sets, so indicators involving a new endpoint remain independent of these events. By contrast, a newly selected matching edge may join two previously used endpoints, so counting new edges alone does not determine how many fresh adjacency constraints the next matching imposes.

Using Boolean Fourier analysis as in~\cite{Wein2021}, together with our normalization and rounding conditions, we obtain an event on which consecutive successful outputs (on $G^{(i)}$ and $G^{(i+1)}$) differ in at most $\eps\Phi/8$ edges. Thus, the number of endpoints outside any fixed vertex set changes by at most $\eps\Phi/4$ per step. Once every adjacency indicator has been resampled after the last matching was selected, the resulting graph is independent of the matchings selected so far. A first-moment bound then ensures that, whp, any large output has more than $\eps\Phi/2$ endpoints outside the union of their endpoint sets. Consequently, the first time this count is at least $\eps\Phi/4$, it is at most $\eps\Phi/2$, and hence lies in the forbidden interval $[\eps\Phi/4,\eps\Phi/2]$; this output becomes the next matching in the forbidden tuple. Our degree and success-probability assumptions ensure that all required events hold simultaneously with positive probability, which yields the contradiction.

\paragraph{Declaration of AI use} GPT-5.5 and 5.6 were used when developing the proofs. See the AI methodology statement at the end of the paper for a detailed account.

\noindent
\textit{Note.} AI use in this paper took place exclusively before NM joined the {Association for Human Mathematics} (AHM).

\subsection{Prior work}
In this section, we discuss related work.

\paragraph{Induced matchings}
Finding a maximum induced matching is NP-hard in general~\cite{cameron1989induced}. For fixed $p$, the size $\beta$ of the largest induced matching in $G(n,p)$ was studied by El Maftouhi and M{\'a}rquez Gordones~\cite{el1994maximum}, who showed that whp $\beta$ lies in an interval of bounded length. Clark~\cite{clark2001strong} later narrowed this down to two values, establishing \emph{two-point concentration}. El Maftouhi~\cite{el1995minimum} also studied maximal induced matchings in $G(n,p)$ for constant $p$, proving that whp every such matching has size at least $(1/2-o(1))\beta$. For the broader range of densities $C/n\le p\le 0.99$, Cooley, Dragani{\'c}, Kang, and Sudakov~\cite{CooleyDraganicKangSudakov2021} determined the size of the largest induced matching up to an arbitrarily small relative error, with the constant $C$ depending on the desired accuracy. These works provide the \emph{statistical threshold} that serves as our benchmark.

Several works have studied algorithms for induced matchings in random regular graphs. By analyzing a random greedy algorithm, Duckworth, Wormald, and Zito~\cite{duckworth2002maximum} obtained constructive lower bounds on the size of the largest induced matching in random cubic graphs. Beis, Duckworth, and Zito~\cite{beis2005large,beis2008packing} studied the more general setting of $k$-separated matchings in random $d$-regular graphs with fixed $d\ge 3$, where $k=1$ corresponds to ordinary matchings and $k=2$ to induced matchings, and obtained constructive lower bounds by analyzing greedy algorithms, together with first-moment upper bounds. These works analyze greedy dynamics using the differential equation method. By contrast, our analysis of greedy requires no tracking of its dynamics: it relies only on the maximality of its output.

\paragraph{Algorithmic gaps for independent sets}
Closely related to our work is the largest independent set problem in random graphs. For $G(n,d/n)$, the optimum is asymptotically $2n\log d/d$ in the double limit $n\to\infty$ followed by $d\to\infty$~\cite{frieze1990independence,frieze1992independence}, while the best known polynomial-time algorithms achieve only half this leading-order size, $n\log d/d$~\cite{lauer2007large}. Using the overlap geometry of pairs of large independent sets, Gamarnik and Sudan~\cite{gamarnik2014limits,gamarnik2017limits} ruled out local algorithms on random $d$-regular graphs above the asymptotic threshold $(1+1/\sqrt{2})n\log d/d$. Rahman and Vir{\'a}g~\cite{rahman2017local} improved this to the optimal threshold $(1+\eps)n\log d/d$ for both random regular and sparse \ER graphs by considering the overlap geometry of many large independent sets, an approach based on the multi-overlap gap property. Matching low-degree lower bounds were later obtained by Wein~\cite{Wein2021} (see also the earlier work of Gamarnik, Jagannath, and Wein~\cite{gamarnik2020low}); our low-degree proof builds on this framework.

These results do not apply directly to $L(G)^2$, whose distribution differs from both the \ER and random regular models. In particular, the edges of $L(G)^2$ inherit strong dependencies from $G$, so the reduction to independent sets does not transfer these hardness bounds to induced matchings.

\paragraph{Online algorithms for random graphs}
For the largest independent set problem in \ER random graphs, Gamarnik, K{\i}z{\i}lda\u{g}, and Warnke~\cite{GamarnikKizildagWarnke2025} established sharp half-optimality bounds for online algorithms over a broad range of edge densities. Their model permits adaptive vertex exposure analogous to ours. Their proofs combine stopping-time arguments with temporally evolving interpolation paths. Related sharp thresholds have been obtained for balanced independent sets in dense bipartite graphs~\cite{dhawan2025sharp}, for independent sets and balanced independent sets in dense hypergraphs~\cite{dhawan2026algorithmic}, and for common induced subgraphs of two independent random graphs~\cite{GamarnikRaczSchoenbach2026}. These works develop overlap-geometry arguments adapted to the online model.

\paragraph{Query complexity}
A complementary line of work allows unrestricted computation but restricts the total number of adjacency queries. In this framework, Feige, Gamarnik, Neeman, R{\'a}cz, and Tetali~\cite{FeigeGamarnikNeemanRaczTetali2020} showed that on $G(n,1/2)$ one can construct, whp, an independent set of size $\alpha\log_2 n$ for some $\alpha>1$ using only $O(n)$ queries. They also proved that for every fixed $0<\delta<2$ and every fixed number of rounds, there exists $\alpha<2$ such that any algorithm making at most $n^\delta$ queries over these rounds fails whp to find an independent set of size $\alpha\log_2 n$. Follow-up work by Cs{\'o}ka and Pongr{\'a}cz~\cite{CsokaPongracz2026} strengthened the multi-round bounds and extended the analysis to finding dense subgraphs. These models impose a global query budget but allow the final output to be chosen after all queries are made. By contrast, our online framework imposes no query budget on the inspected part of the graph but requires irrevocable decisions.

\subsection{Organization}

Section~\ref{sec:models-results} presents our main results about online and low-degree algorithms. Section~\ref{sec:greedy} proves the sharp bounds for online algorithms, including the guarantee for the greedy algorithm and the matching impossibility result.
Sections~\ref{sec:positive-low-degree} and~\ref{sec:low-degree} prove the low-degree achievability and hardness results respectively.

\section{Models and results}
\label{sec:models-results}

\subsection{Online achievability}
\label{subsec:greedy-results}

We begin with a natural greedy algorithm for finding a large induced matching, which belongs to a class of online algorithms to be introduced in Section~\ref{subsec:online-barrier}. 
Fix an arbitrary ordering $1, \dots, n$ of the vertices.
The Greedy algorithm proceeds in rounds $t = 1, \dots, n$: in round $t$, vertex $t$ is inspected and its adjacencies to vertices $1, \dots, t-1$ are revealed.
Let $M_t \subseteq E(G[t])$ denote the induced matching after round $t$, where $G[t]$ is the subgraph of $G$ induced by the first $t$ vertices.
We also maintain a \emph{buffer} $B_t$, consisting of the vertices in $[t] \setminus V(M_t)$ with no neighbors in $V(M_t)$; by construction, $B_t$ is an independent set.
Initially, $M_0 = B_0 = \emptyset$.
In round $t$, the algorithm proceeds as follows:
\begin{itemize}
    \item If vertex $t$ has no neighbors in $V(M_{t-1}) \cup B_{t-1}$, set $M_t = M_{t-1}$ and $B_t = B_{t-1} \cup \{t\}$.
    \item If vertex $t$ has no neighbors in $V(M_{t-1})$ but has a neighbor in $B_{t-1}$, choose an arbitrary $u_t \in N(t) \cap B_{t-1}$, and set $M_t = M_{t-1} \cup \{\{u_t,t\}\}$ and $B_t = B_{t-1} \setminus N(t)$.
    \item Otherwise (i.e., if $t$ has a neighbor in $V(M_{t-1})$), set $M_t = M_{t-1}$ and $B_t = B_{t-1}$.
\end{itemize}
The output is the matching $M_n$; a formal description is given in Algorithm~\ref{alg:greedy-IM}.
The following theorem shows that, whp, Greedy finds an induced matching of size at least roughly half the maximum.

\begin{theorem}
\label{thm:greedy-main}
Fix $\kappa\in(0,1)$ and $\eps \in (0,1/2)$, and let $L$ be given by~\eqref{eq: def of b and L}.
There is a constant $C=C(\eps,\kappa) > 0$ such that, for all sufficiently large $n$ and all $C/n\le p\le1-\kappa$, the output $M_n$ of Greedy satisfies
\[
  \Prb\left[ |M_n| \ge 
    \left(\frac12-\eps\right)L \right]
  \ge  
  1-\exp\left(-n^\eps\right).
\]
\end{theorem}

The proof is deferred to Section~\ref{sec:greedy-analysis}.
The output $M_n$ of Greedy is \emph{maximal}: no edge can be added to $M_n$ while preserving the induced matching property.
In fact, Proposition~\ref{prop:maximal-matching-half-optimal} shows that, whp, \emph{every} maximal induced matching is at least half-optimal, which immediately implies Theorem~\ref{thm:greedy-main}. This observation lets us avoid the more involved coupling and stopping-time arguments typically used to analyze greedy algorithms on random graphs (see, e.g.,~\cite{grimmett1975colouring, dhawan2025sharp, dhawan2026algorithmic, GamarnikKizildagWarnke2025}).

\subsection{Online barrier}
\label{subsec:online-barrier}

We now consider a broad class of online algorithms and show that, for every fixed $\eps>0$, no algorithm in this class finds an induced matching of size larger than $(1/2+\eps)L$, except with vanishing probability.
Since Greedy belongs to this class, the guarantee in Theorem~\ref{thm:greedy-main} is optimal.

\begin{definition}[Online algorithm]
\label{def:online-algorithm}
An online algorithm maintains, after each round $t$, a set $S_t$ of inspected vertices and an induced matching $M_t \subseteq E(G[S_t])$.
Initially, $M_0=S_0=\emptyset$.
In round $t$, the algorithm either terminates or proceeds as follows:
\begin{enumerate}
\item The algorithm selects a vertex $x_t \in [n]\setminus S_{t-1}$ and inspects it, revealing its adjacencies to the vertices in $S_{t-1}$. Set $S_t=S_{t-1}\cup\{x_t\}$.

\item The algorithm may accept one edge $\{x_t,u_t\}$, where $u_t \in S_{t-1}\setminus V(M_{t-1})$ is a neighbor of $x_t$, and set $M_t=M_{t-1}\cup\{\{x_t,u_t\}\}$.
This update is allowed only if $M_t$ is an induced matching.
If no edge is accepted, set $M_t=M_{t-1}$.
\end{enumerate}
Accepted edges cannot be removed.
If the algorithm has not terminated earlier, it terminates after round $n$.
Its output $\Mout$ is the induced matching held upon termination.
\end{definition}

Algorithms in this class are online in the following sense.
First, new adjacency information is revealed only in Step~1 of each round.
Second, every choice---including the inspection order, whether and which edge to accept, and when to terminate---is a measurable function of all previous choices and revealed adjacencies, together with auxiliary randomness independent of the graph.
There is no restriction on the algorithm's memory or computation.
The following result establishes the half-optimality barrier for this class.

\begin{theorem}
\label{thm:broad-main}
Fix $\kappa\in(0,1)$ and $\eps>0$.
There is a constant $C=C(\eps,\kappa)>0$ such that, for all sufficiently large $n$, all $C/n\le p\le1-\kappa$, and every algorithm as in Definition~\ref{def:online-algorithm},
\begin{equation*}
  \Prb\left[
    |\Mout|\le\left(\frac12+\eps\right)L
  \right]
  \ge 1 - \exp\left(-\frac{\eps^2}{2}L\,\log (np)\right).
\end{equation*}
\end{theorem}

The proof is given in Section~\ref{sec:online-upper-proof}.
Together, Theorems~\ref{thm:greedy-main} and~\ref{thm:broad-main} show that $(1/2+o(1))L$ is the optimal size attainable by online algorithms, and that Greedy attains it.

\subsection{Low-degree achievability}
\label{subsec:low-degree-results}
\label{subsec:low-degree-achievability}

We now turn to low-degree polynomial algorithms for sparse random graphs \(G(n,d/n)\), beginning with achievability. Throughout, we first choose $d>1$ sufficiently large and then let $n\to\infty$ with $d$ fixed. 
Let
\[
    \mathcal E_n=\binom{[n]}2
    \qquad\text{and}\qquad
    m=|\mathcal E_n|=\binom n2.
\]
We identify a graph $G$ on $[n]$ with its vector of edge indicators
\[
    (G_e)_{e\in\mathcal E_n}\in\{0,1\}^m,
    \qquad\text{where}\qquad G_e=\bbone_{\{e\in E(G)\}}.
\]
A polynomial $f(G)$ of degree at most $D$ refers to a map
\[
    f=(f_e)_{e\in\mathcal E_n}:\RR^m\longrightarrow\RR^m
\]
such that each coordinate $f_e(G)$ is a polynomial of degree at most $D$ in the $m$ input variables $(G_{e'})_{e'\in\mathcal E_n}$. 
We say that $f(G;\xi)=(f_e(G;\xi))_{e\in\mathcal E_n}$ is a \emph{random} polynomial of degree at most $D$ if its coefficients depend on a random seed $\xi$, sampled independently of $G$, and if it has degree at most $D$ for every fixed $\xi$.
Fixing the seed $\xi$ gives a \emph{deterministic} polynomial. 
We suppress the notation $\xi$ when clear from the context.

We now specify the procedure used to map a polynomial $f$ on $G$ to an induced matching.
\begin{definition}\label{def:rounded-polynomial-output}
    Let $G$ be a graph on $[n]$, and let $f : \RR^m \to \RR^m$ be a polynomial. 
    We say that two distinct edges of $G$ \emph{conflict} if they share an endpoint or an edge of $G$ joins an endpoint of one to an endpoint of the other (equivalently, if they are adjacent in the square of the line graph of $G$).
    Let
    \begin{align*}
        A &= \inbraces{e \in E(G) : f_e(G) \geq 1}, \\
        \tilde{A} &= \inbraces{e \in A :
        \text{there is no distinct $e'\in A$ that conflicts with $e$ in $G$}}, \\
        B &= \inbraces{e \in \mathcal E_n : f_e(G)  \in \inparen{\frac{1}{2},1}}.
    \end{align*}
    For $\eta > 0$, the output induced matching is 
    \begin{align*}
        \mathcal M_f^\eta(G) = \begin{cases}
            \tilde{A}, & \text{ if }\abs{A \setminus \tilde{A}} + \abs{B} \leq \eta m \\
            \varnothing, & \text{ otherwise}.
        \end{cases}
    \end{align*}
\end{definition}

\medskip
\noindent
In the above definition, $\eta$ is a tolerance parameter for the rounding error, $A$ is the candidate edge set, $\tilde A$ is the subset surviving conflict pruning, and $B$ is the set of edges about which $f$ is unsure.

Next, we define a performance metric for a random polynomial.
\begin{definition}\label{def:k_delta_gamma_eta_optimizes}
    Let $k > 0$, $\delta \in [0,1]$, $\gamma \geq 1$, $\eta > 0$ be parameters. We say that a random polynomial $f : \RR^m \to \RR^m$ is a $(k,\delta,\gamma,\eta)$-optimizer for the induced matching problem in $G(n,d/n)$ if the following are satisfied for $G \sim G(n,d/n)$, jointly over $G$ and the independent seed:
    \begin{itemize}
        \item $\E_{G,f}\left[\norm{f(G)}^2\right] \leq \gamma k$,
        \item $\Prb_{G,f} \insquare{\abs{\mathcal M_f^\eta(G)} \geq k} \geq 1 - \delta$.
    \end{itemize}
\end{definition}

\medskip
\noindent
In other words, a $(k,\delta,\gamma,\eta)$-optimizer $f$ yields an induced matching $\mathcal M_f^\eta(G)$ with at least $k$ edges with probability at least $1-\delta$, over both the graph and the random seed; 
in addition, the expected squared norm of $f(G)$ is limited by $\gamma k$, allowing us to control how much the polynomial output changes when the graph changes, as needed for the hardness arguments later.
We remark that Definitions~\ref{def:rounded-polynomial-output} and~\ref{def:k_delta_gamma_eta_optimizes} are analogous to the corresponding definitions for independent sets in \cite{Wein2021}.

In the sparse regime, it is more convenient to work with the threshold $\Phi:=(n\log d)/d$. Note that for fixed $d>1$, the relation to the previous threshold $L$ in the online setting is 
\begin{equation*}
  L
  =\frac{\log d}{-\log(1-d/n)}
  =\Phi-\frac12\log d+O_d(n^{-1})
  =\Phi+O_d(1).
\end{equation*} 
Our main positive result shows that there is a constant-degree polynomial that finds an induced matching with at least $(1/2-\eps)\Phi$ edges whp.
\begin{theorem}
\label{thm:sparse-positive-low-degree}
For every $0<\eps<1/2$ and every sufficiently large fixed $d$, there exists an integer $D=D(d,\eps)\geq1$ and constants $\gamma=\gamma(d,\eps)\geq1$ and $c=c(d,\eps)>0$ such that, for all sufficiently large $n$, there is a deterministic polynomial of degree at most $D$ that is a
\begin{equation}
    \left(
        \left(\frac12-\eps\right)\Phi,
        \exp(-cn^{1/3}),
        \gamma,
        \frac{\eps\Phi}{128m}
    \right)\text{-optimizer}
    \label{eq:positive-optimizer-parameters}
\end{equation}
for the induced matching problem in $G(n,d/n)$.  In particular, this low-degree algorithm outputs an induced matching with at least $(1/2-\eps)\Phi$ edges with probability $1-\exp(-cn^{1/3})$.
\end{theorem}

As alluded to in Section~\ref{subsec:proof-techniques}, we construct a local algorithm based on finding independent sets in the \emph{conflict graph} $\mathrm{Line}(G)^2$ and approximate its decisions by a constant-degree polynomial. 
The details of the algorithm and its proof are given in Section~\ref{sec:positive-low-degree}.

\subsection{Low-degree barrier}
\label{subsec:low-degree-hardness}

We now give the corresponding hardness result for low-degree polynomial algorithms in the sparse graph $G(n,d/n)$ setting. We continue to use the notation and definitions introduced in Section~\ref{subsec:low-degree-achievability}. 
\begin{theorem}
\label{thm:sparse_low_degree_hardness}
    For every $\eps>0$ and every sufficiently large fixed $d$, there are constants $c_1=c_1(\eps,d)>0$ and $c_2=c_2(\eps,d)>0$ for which the following holds for all sufficiently large $n$.  
    Let $D=D_n\in\NN$, $\gamma=\gamma_n\geq1$, $\delta=\delta_n\in[0,1]$, and $\eta=\eta_n>0$ satisfy
    \[
        D \leq \frac{c_1 n}{\gamma \log n},
        \qquad
        \delta \leq \exp(-c_2\gamma D \log n),
        \qquad\text{and}\qquad
        \eta\leq \frac{\eps\Phi}{64m},
    \]
    and let $k = \inparen{\frac{1}{2} + \eps} \Phi$. 
    Then there does not exist a random polynomial of degree at most-$D$ that is a $(k,\delta,\gamma,\eta)$-optimizer for the induced matching problem in $G(n,d/n)$.
\end{theorem}

The proof builds upon the overlap gap property approach in \cite{Wein2021} and is given in Section~\ref{sec:low-degree}.
We note that the optimizer in Theorem \ref{thm:sparse-positive-low-degree} is within the parameter range of Theorem~\ref{thm:sparse_low_degree_hardness} for sufficiently large $n$. Indeed, the constants $D$ and $\gamma$ in Theorem \ref{thm:sparse-positive-low-degree} are fixed in $n$, and $e^{-cn^{1/3}}$ is eventually smaller than any fixed inverse power of $n$.
    For these fixed values of $D$ and $\gamma$, the bound on $\delta$ in Theorem~\ref{thm:sparse_low_degree_hardness} becomes $\delta\leq\exp(-c_2\gamma D\log n)=n^{-c_2\gamma D}$.
    Together, the two theorems establish a sharp half-optimality threshold for low-degree optimization: constant degree suffices to approach $\Phi/2$, while the hardness result rules out going above the leading constant $1/2$, even for degrees growing with $n$, at the specified success probabilities.

\section{Sharp bounds for online algorithms}
\label{sec:greedy}

In this section, we prove our main results on online algorithms. In Section~\ref{sec:greedy-analysis}, we prove the achievability result, Theorem~\ref{thm:greedy-main}, and in Section~\ref{sec:online-upper-proof}, we prove the impossibility result, Theorem~\ref{thm:broad-main}.
Throughout, we use the parameters from~\eqref{eq: def of b and L}:
\[
    b\coloneqq \frac{1}{1-p},\qquad
    L\coloneqq \log_b(np).
\]

\subsection{Achievability by the greedy algorithm} \label{sec:greedy-analysis}

In this subsection, we prove Theorem~\ref{thm:greedy-main}.
The Greedy algorithm is formally described below.

\begin{breakablealgorithm}
\caption{Greedy algorithm for building an induced matching in $G(n,p)$}
\label{alg:greedy-IM}
\begin{algorithmic}[1]
\State $M_0 \gets \emptyset$, \qquad $B_0 \gets \emptyset$
\For{$t \gets 1, \ldots, n$}
  \State Inspect vertex $t$ and reveal its adjacencies to vertices $1, \ldots, t-1$.
  \If{$N(t) \cap V(M_{t-1}) = \emptyset$ and $N(t) \cap B_{t-1} = \emptyset$}
    \State $M_t \gets M_{t-1}$, \qquad $B_t \gets B_{t-1} \cup \{t\}$.
  \ElsIf{$N(t) \cap V(M_{t-1}) = \emptyset$ and $N(t) \cap B_{t-1} \neq \emptyset$}
    \State Let $u_t \in N(t) \cap B_{t-1}$ be arbitrary.
    \State $M_t \gets M_{t-1} \cup \{\{u_t,t\}\}$, \qquad $B_t \gets B_{t-1} \setminus N(t)$.
  \Else
    \State $M_t \gets M_{t-1}$, \qquad $B_t \gets B_{t-1}$.
  \EndIf
\EndFor
\State \Return $M_{n}$.
\end{algorithmic}
\end{breakablealgorithm}

It is easy to verify that $M_n$ is a maximal induced matching. Indeed, an edge $xy$ can be added to $M_n$ while keeping it induced only if $x, y \notin V(M_n)$ and neither $x$ nor $y$ has a neighbor in $V(M_n)$, that is, only if $x, y \in B_n$; but $B_n$ is an independent set.
Theorem~\ref{thm:greedy-main} therefore follows from the next proposition, which shows that, with high probability, every maximal induced matching of $G(n,p)$ is half-optimal.

\begin{proposition} \label{prop:maximal-matching-half-optimal}
    Fix $\kappa \in (0,1)$ and $\eps \in (0, 1/2)$, and let $b$ and $L$ be as in~\eqref{eq: def of b and L}. There exists $C = C(\eps,\kappa) > 1$ such that, for all sufficiently large $n$ and uniformly over $C/n \le p \le 1-\kappa$,
    \begin{align*}
        \Prb\left[
        \text{there is a maximal induced matching $M$ with }
        |M| < \left(\frac12-\eps\right)L
      \right] &\le L\exp\left(- \frac{(np)^{2\eps}}{30p}\right) \\
      &= \exp\left(-\Omega\left(n^\eps\right)\right).
    \end{align*}
    In particular, with high probability every maximal induced matching has at least $(1/2-\eps)L$ edges.
\end{proposition}

\begin{proof}
    Let $G \sim G(n, p)$. The case $k=0$ is trivial, since the empty matching is maximal only if $G$ has no edges, which occurs with probability $(1-p)^{\binom n2}$.
    Fix $1\le k < \left(\frac12-\eps\right)L$ and a set $S \subseteq [n]$ with $|S| = 2k$.
    Let $R_S \subseteq [n] \setminus S$ be the random set of vertices with no neighbors in $S$.
    Let $\cA$ denote the event that $G[S]$ is a perfect matching, and let $\cB$ denote the event that $R_S$ is an independent set.
    Since an induced matching with vertex set $S$ is maximal if and only if $R_S$ is independent, we have
    \begin{equation}\label{eq: AB decomp}
        \Prb[S \text{ spans a maximal induced matching}] = \Prb[\cA]\Prb[\cB \mid \cA].
    \end{equation}
    We bound each factor separately.
    First, since $S$ admits $(2k)!/(2^kk!)$ perfect matchings,
    \begin{equation}\label{eq: A bound}
        \Prb[\cA] = \frac{(2k)!}{2^kk!}(1 - p)^{\binom{2k}{2} - k}p^k.
    \end{equation}
    Next, $\cA$ depends only on the pairs inside $S$, $R_S$ only on the pairs between $S$ and $[n]\setminus S$, and $\cB$ given $R_S$ only on the pairs inside $R_S$. Hence
    \begin{equation}\label{eq: B reduction}
        \Prb[\cB \mid \cA] = \E\left[(1-p)^{\binom{|R_S|}{2}}\right].
    \end{equation}
    Since $|R_S| \sim \Bin\left(n-2k, (1-p)^{2k}\right)$ and $(1-p)^{2k} \ge (1-p)^{(1-2\eps)L} = (np)^{-(1-2\eps)}$, we have
    \[
        \E[|R_S|] = (1-p)^{2k}(n-2k) \geq \frac{n - 2k}{(np)^{1 - 2\eps}} \geq \frac{(np)^{2\eps}}{2p} \eqqcolon \mu,
    \]
    where the last step uses $2k \le L \le n/2$, which holds throughout the range of $p$ under consideration.
    By the Chernoff bound,
    \[
        \Prb[|R_S| \leq \mu/2] \leq \exp\left(-\frac{\mu}{8}\right).
    \]
    Since $\mu \ge 4$ and $\mu\log b \ge \mu p = (np)^{2\eps}/2 \ge 2$ for $C$ sufficiently large, it follows that
    \[
        \E\left[(1-p)^{\binom{|R_S|}{2}}\right] \leq \exp\left(-\frac{\mu}{8}\right) + (1- p)^{\binom{\lceil\mu/2\rceil}{2}} \leq \exp\left(-\frac{\mu}{8}\right) + b^{-\frac{\mu^2}{16}} \leq 2\exp\left(-\frac{(np)^{2\eps}}{20p}\right).
    \]
    Combining \eqref{eq: AB decomp}, \eqref{eq: A bound}, \eqref{eq: B reduction}, and the above, we obtain
    \[
        \Prb[S \text{ spans a maximal induced matching}] \leq \frac{(2k)!}{2^{k-1}k!}(1 - p)^{\binom{2k}{2} - k}p^k\exp\left(-\frac{(np)^{2\eps}}{20p}\right).
    \]
    Using $\binom{2k}{2} - k = 2k(k-1)$, $\binom{n}{2k}(2k)!\le n^{2k}$, and $k! \ge (k/e)^k$, it follows that the expected number of maximal induced matchings of size $k$ is at most
    \begin{align*}
        &\binom{n}{2k}\frac{(2k)!}{2^{k-1}k!}(1 - p)^{2k(k-1)}p^k\exp\left(-\frac{(np)^{2\eps}}{20p}\right) \\
        &\qquad \leq 2\,n^{2k}\left(\frac{e}{2k}\right)^k(1 - p)^{2k(k-1)}p^k\exp\left(-\frac{(np)^{2\eps}}{20p}\right) \\
        &\qquad = 2\exp\left(2k\log n + k \log \left(\frac{ep}{2k}\right) - 2k(k-1)\log b - \frac{(np)^{2\eps}}{20p}\right) \\
        &\qquad = 2\exp\left(2k\log (np) + k \log \left(\frac{e}{2pk}\right) - 2k(k-1)\log b - \frac{(np)^{2\eps}}{20p}\right) \\
        &\qquad \leq \exp\left(2k(L - k + 1)\log b + \frac{1}{2p} + 1 - \frac{(np)^{2\eps}}{20p}\right),
    \end{align*}
    where in the last step we used $\log (np) = L \log b$, $\log 2\le 1$, and the inequality $x\log (1/x) \le 1/e$ with $x = 2pk/e$.
    Since $2k(L-k+1)$ is increasing in $k$ for $k \le (L+1)/2$, and $\frac{(np)^{2\eps}}{20p} - \frac{1}{2p} - 1 \ge \frac{(np)^{2\eps}}{25p}$ for $C$ sufficiently large, Markov's inequality gives
    \begin{align*}
        &\Prb[G \text{ contains a maximal induced matching of size } k] \\
        &\qquad \leq \exp\left((1 - 2\eps)L\left(\left(\frac{1}{2} + \eps\right)L + 1\right)\log b - \frac{(np)^{2\eps}}{25p}\right) \\
        &\qquad \leq \exp\left(\left(\frac{(1 - 4\eps^2)L^2}{2} + L\right)\log b - \frac{(np)^{2\eps}}{25p}\right) \\
        &\qquad \leq \exp\left(\frac{(1 - 3\eps^2)L^2\log b}{2} - \frac{(np)^{2\eps}}{25p}\right),
    \end{align*}
    where the last step uses $L \ge 2/\eps^2$, which holds for $C$ sufficiently large.
    Since $\log b \ge p$, we have $L^2 \log b = \frac{\log^2(np)}{\log b} \leq \frac{\log^2(np)}{p} \ll \frac{(np)^{2\eps}}{p}$ for $C$ sufficiently large, and hence
    \[
        \Prb[G \text{ contains a maximal induced matching of size } k] \leq \exp\left(- \frac{(np)^{2\eps}}{30p}\right).
    \]
    Finally, since $p\le 1$ and $2\eps-1<0$, we have $(np)^{2\eps}/p = n^{2\eps}p^{2\eps-1} \ge n^{2\eps}$. Taking a union bound over the at most $L\le n$ possible values of $k$, we conclude that, for $n$ sufficiently large,
    \begin{align*}
        \Prb\left[G \text{ contains a maximal induced matching of size } < \left(\frac{1}{2} - \eps\right)L\right] &\leq L\exp\left(- \frac{(np)^{2\eps}}{30p}\right) \\
        &\le n\exp\left(-\frac{n^{2\eps}}{30}\right) \le \exp\left(-n^\eps\right),
    \end{align*}
    completing the proof.
\end{proof}

\subsection{Impossibility above the computational threshold}
\label{sec:online-upper-proof}

In this subsection, we prove Theorem~\ref{thm:broad-main}; that is, we show that every online algorithm in the sense of Definition~\ref{def:online-algorithm} constructs an induced matching of size at least $(1/2+\eps)L$ with probability at most $\exp\bigl(-\frac{\eps^2}{2}L\log(np)\bigr)$.
To this end, we define
\begin{equation}\label{eq: r and s}
    m^* \coloneqq \left(\frac{1}{2} + \eps\right)L, \qquad r \coloneqq \left(\frac{1}{2} + \frac{\eps}{2}\right)L, \qquad s \coloneqq m^* - r = \frac{\eps L}{2}.
\end{equation}
For simplicity, we ignore floors and ceilings and treat $m^*$, $r$, and $s$ as integers.

Let $\cA$ be an online algorithm in the sense of Definition~\ref{def:online-algorithm}.
By a standard averaging argument (see~\cite[Section~4.1]{GamarnikKizildagWarnke2025}), we may assume that $\cA$ is deterministic, i.e., that the only randomness comes from the input graph $G$ and not from the algorithm's internal randomness.
Let $\cA_t(G)$ denote the induced matching constructed after the first $t$ vertices have been exposed, with $\cA_t(G)\coloneqq\Mout$ if the algorithm has terminated by round $t$; in particular, $\cA_n(G)=\Mout$. Let $\tau_r$ be the first time at which this matching has size at least $r$:
\begin{equation}\label{eq: def tau r}
    \tau_r \coloneqq \min\{t \in [n]\,:\, |\cA_t(G)| \ge r\},
\end{equation}
with the convention that $\min\emptyset = \infty$.
Since every edge accepted in round $t$ contains the vertex exposed in round $t$, at most one edge is accepted per round; in particular, $|\cA_{\tau_r}(G)| = r$ whenever $\tau_r < \infty$.
The following lemma bounds the probability that the algorithm accepts $s$ further edges after time $\tau_r$.

\begin{lemma}\label{lemma: new s edges}
    For every $t \in [n]$ with $\Prb[\tau_r = t] > 0$,
    \[
        \Prb[\text{at least $s$ edges are accepted after time $t$} \mid \tau_r = t] \le \binom{n}{s}(1-p)^{2rs + \binom{s}{2}}.
    \]
\end{lemma}

\begin{proof}
    Let $M^* \coloneqq \cA_{t}(G)$, so that $|V(M^*)| = 2r$ on the event $\{\tau_r = t\}$.
    We claim that if $s$ further edges are accepted, then there is a set $X$ of $s$ vertices, none of which has been exposed by time $t$, such that (i) $X$ is an independent set, and (ii) no edge joins a vertex of $X$ to a vertex of $V(M^*)$.
    Indeed, let $e_1, \ldots, e_s$ be edges accepted after time $t$, and for each $i$ let $x_i$ be the endpoint of $e_i$ exposed in the round in which $e_i$ is accepted. By Definition~\ref{def:online-algorithm}, this round is later than $t$, so $x_i$ is unexposed at time $t$.
    Let $X \coloneqq \{x_1, \ldots, x_s\}$; these vertices are distinct, since $e_1, \ldots, e_s$ are pairwise disjoint.
    Since accepted edges are never removed, the final output contains $M^*\cup\{e_1,\dots,e_s\}$. Thus $X$ must be independent, as otherwise $\{e_1, \ldots, e_s\}$ would not be an induced matching.
    Furthermore, if $ux_i \in E(G)$ for some $u \in V(M^*)$ and $x_i \in X$, then $M^* \cup \{e_i\}$ would not be an induced matching.

    Now condition on the information revealed by time $t$. Since $\cA$ is deterministic, this information determines whether $\tau_r = t$, as well as $M^*$ and the set of unexposed vertices.
    The pairs within the set of unexposed vertices, and between this set and $V(M^*)$, have not been revealed, so their edge indicators are independent of this information.
    Hence, for any fixed set $X$ of $s$ unexposed vertices, properties (i) and (ii) hold with conditional probability $(1-p)^{2rs + \binom{s}{2}}$.
    Taking a union bound over the at most $\binom{n}{s}$ choices of $X$, and then averaging over the information revealed by time $t$ on the event $\{\tau_r=t\}$, completes the proof.
\end{proof}

We are now ready to prove Theorem~\ref{thm:broad-main}.
Since $m^*\ge r$, and since $|\cA_n(G)|\ge m^*$ together with $\tau_r=t$ implies that at least $m^*-r=s$ edges are accepted after time $t$, Lemma~\ref{lemma: new s edges} gives
\begin{align}
    \Prb[|\cA_n(G)| \ge m^*] &= \Prb[\tau_r < \infty,\, |\cA_n(G)| \ge m^*] \nonumber\\
    &= \sum_{t = 1}^n\Prb[|\cA_n(G)| \ge m^* \mid \tau_r = t]\,\Prb[\tau_r = t] \nonumber\\
    &\le \sum_{t = 1}^n\Prb[\text{at least $s$ edges are accepted after time $t$} \mid \tau_r = t]\,\Prb[\tau_r = t] \nonumber\\
    &\le \binom{n}{s}(1 - p)^{2rs + \binom{s}{2}}, \label{eq: ub on prob in terms of r and s}
\end{align}
where the sums range over those $t$ with $\Prb[\tau_r = t] > 0$.

Before bounding~\eqref{eq: ub on prob in terms of r and s}, we record an inequality that will be used below.
Note that
\[
    \frac{n}{L} = \frac{n}{\log_b(np)} = \frac{np}{\log (np)}\cdot\frac{\log b}{p}.
\]
Since the function $f(x) = -\frac{\log(1-x)}{x}$ is increasing on $(0, 1)$ and $C/n \le p \le 1 - \kappa$, for $C$ sufficiently large in terms of $\kappa$ we have
\begin{equation}\label{eq: n over L}
    \frac{n}{L} \le \frac{np}{\log C}\cdot\frac{\log (1/\kappa)}{1 - \kappa} \le np.
\end{equation}

We now bound~\eqref{eq: ub on prob in terms of r and s}.
Using $\binom{n}{s} \le (en/s)^s$, we obtain
\begin{align}
    \binom{n}{s}(1 - p)^{2rs + \binom{s}{2}} &= \exp\left(\log \binom{n}{s} - \left(2rs + \binom{s}{2}\right)\log b\right) \nonumber \\
    &\le \exp\left(s\log \left(\frac{en}{s}\right) - \left(2rs + \binom{s}{2}\right)\log b\right) \nonumber \\
    &= \exp\left(s\left(1 + \log \left(\frac{n}{s}\right) - \left(2r + \frac{s-1}{2}\right)\log b\right)\right). \label{eq: exp bound in terms of s}
\end{align}
By~\eqref{eq: r and s}, we have $2r + \frac{s-1}{2} = \left(1 + \frac{5\eps}{4}\right)L - \frac{1}{2}$ and $\frac{n}{s} = \frac{2n}{\eps L}$.
Hence, using~\eqref{eq: n over L} and $L\log b = \log(np)$, we get
\begin{align*}
    1 + \log \left(\frac{n}{s}\right) - \left(2r + \frac{s-1}{2}\right)\log b &= 1 + \log \left(\frac{2n}{\eps L}\right) - \left(\left(1+ \frac{5\eps}{4}\right)L - \frac{1}{2}\right)\log b \\
    &\le 1 + \log\left(\frac{2}{\eps}\right) + \log (np) - \left(1+ \frac{5\eps}{4}\right)\log(np) + \frac{\log b}{2} \\
    &= 1 + \log\left(\frac{2}{\eps}\right) + \frac{\log b}{2} - \frac{5\eps}{4}\log (np) \\
    &\le -\eps\log(np),
\end{align*}
where the last step uses $\log b \le \log(1/\kappa)$ and $np \ge C$, and holds for $C$ sufficiently large in terms of $\eps$ and $\kappa$.
Plugging this into~\eqref{eq: exp bound in terms of s}, and recalling that $s = \eps L/2$, we conclude that
\[
    \Prb[|\cA_n(G)| \ge m^*] \le \exp\left(-\eps s \log (np)\right) = \exp\left(-\frac{\eps^2}{2}L\log (np)\right),
\]
which completes the proof of Theorem~\ref{thm:broad-main}.

\section{Low-degree achievability: proof of Theorem~\ref{thm:sparse-positive-low-degree}}
\label{sec:positive-low-degree} %

The purpose of this section is to prove Theorem~\ref{thm:sparse-positive-low-degree}, our positive result in the sparse regime.

We begin by describing the algorithm. Recall that for a graph $G$, the line graph $\mathrm{Line}(G)$ is the graph whose vertices are the edges of $G$, in which two vertices are adjacent whenever the corresponding edges of $G$ share an endpoint. For a graph $G$, define its \emph{conflict graph} $H(G)$ by
\begin{equation}
    H(G)=\mathrm{Line}(G)^2.
    \label{eq:IM-conflict-IS}
\end{equation}
The induced matchings of $G$ are precisely the independent sets of $H(G)$. %

For an edge $e$ of $G$ and an integer $r\ge 0$, define the \emph{closed ball} of radius $r$ about $e$ in $\mathrm{Line}(G)$ by
\[
    \mathsf B_r(G,e)
    \coloneqq \{a\in E(G):
           \mathrm{dist}_{\mathrm{Line}(G)}(e,a)\leq r\}.
\]
We view $\mathsf B_r(G,e)$ as the subgraph of $G$ consisting of these edges and their endpoints, rooted at the edge $e$.

Theorem~\ref{thm:sparse-positive-low-degree} follows from two propositions. The first shows that a local rule, Algorithm~\ref{alg:finite-color-conflict}, produces an induced matching of size at least $(1/2-\eps/2)\Phi$, where we recall that $\Phi = n\log d / d$. The second shows that a low-degree polynomial reproduces the decisions of this rule up to an error of at most $\eps\Phi/128$.

Throughout, $\eps$ and a sufficiently large $d$ are fixed deterministic parameters, chosen before sampling $G$ and held fixed as $n\to\infty$. We take $G\sim G(n,d/n)$. A random $Q$-coloring $\chi=(\chi(a))_{a\in\binom{[n]}2}$ assigns to each potential edge an independent uniform color in $[Q]$, independently of $G$. In other words, we color every potential edge first; once $G$ is sampled, the colors of absent edges are simply ignored. %

\begin{algorithm}[H]
\caption{Local induced matching}
\label{alg:finite-color-conflict}
\begin{algorithmic}[1]
\Require A graph $G$ on $[n]$ and an integer $Q\geq2$
\State Form $H=\mathrm{Line}(G)^2$ on vertex set $E(G)$.
\State Independently assign each potential edge $e\in\binom{[n]}2$
a uniform color $\chi(e)\in[Q]$.
\State Set $C\gets\{e\in V(H):\exists f\sim_H e,\ \chi(f)=\chi(e)\}$
and $I_0\gets\varnothing$.
\For{$j=1,\ldots,Q$}
    \State $I_j\gets I_{j-1}\cup
    \{e\in V(H)\setminus C:\chi(e)=j,\
                    N_H(e)\cap I_{j-1}=\varnothing\}$.
\EndFor
\State \Return $I_Q$, interpreted as an induced matching in $G$.
\end{algorithmic}
\end{algorithm}

\begin{proposition}
\label{prop:positive-local-precursor}
For every $0<\eps<1/2$ and every sufficiently large fixed $d$, there is an integer $Q=Q(d,\eps)\geq2$ with the following properties.
Let $\chi=(\chi(a))_{a\in\binom{[n]}2}$ denote the colors in Algorithm~\ref{alg:finite-color-conflict}, which are independent uniform elements of $[Q]$, sampled independently of $G$.
The output $I_Q=I_Q(G,\chi)$ is always an induced matching, and whether a present edge $e$ belongs to $I_Q$ is determined by the colored ball $\mathsf B_r(G,e)$ with $r=2Q+4$.
Moreover, for some $c_1=c_1(d,\eps)>0$ and all sufficiently large $n$,
\begin{equation}
    \Prb_{G,\chi}\left[
        |I_Q|\geq\left(\frac12-\frac{\eps}{2}\right)\Phi
    \right]
    \geq 1-e^{-c_1n^{1/3}}.
    \label{eq:positive-local-output}
\end{equation}
\end{proposition}

Next, we show that the output of this local algorithm can be approximately reproduced by a low-degree polynomial. An \emph{$r$-edge-local rule} is an isomorphism-invariant $\{0,1\}$-valued function of a marked ball $\mathsf B_r(G,e)$. %
Here the marks $(\xi_a)_{a\in\binom{[n]}{2}}$ are i.i.d.\ with a fixed law on a finite nonempty space $\mathcal X$, independent of $G$. The rule gives decisions $h_e(G,\xi)\in \{0,1\}$ that depend only on $\mathsf B_r(G,e)$ and its marks, that is, $h_e(G,\xi)=h_e(\mathsf B_r(G,e),\xi)$, with $h_e=0$ whenever $e \notin E(G)$. %
Set $M_h(G,\xi)\coloneqq\{e:h_e(G,\xi)=1\}$.

In the next proposition, we use the sets $A$, $\tilde A$, and $B$ associated with a polynomial $f$ of the edge indicators, as in Definition~\ref{def:rounded-polynomial-output}.

\begin{proposition}
\label{prop:positive-edge-local-to-polynomial}
Fix $d>1$, an integer $r\geq1$, and $\theta>0$. Suppose $h$ is an $r$-edge-local rule such that $M_h(G,\xi)$ is an induced matching for every graph $G$ and every mark realization $\xi$. Then there exist an integer $D\geq1$ and constants $\Gamma,c_2>0$, depending only on $h$ and $(d,r,\theta)$, such that, for all sufficiently large $n$, there is a family of polynomials $f(\,\cdot\,;\xi)$ of degree at most $D$ satisfying
\begin{equation}
    \sup_\xi\E_G\|f(G;\xi)\|^2\leq\Gamma n.
    \label{eq:positive-conversion-norm}
\end{equation}
Moreover, with $\eta=\theta n/m$ and the sets $A,\tilde A,B$ as in Definition~\ref{def:rounded-polynomial-output},
\begin{equation}
    \Prb_{G,\xi}\left[
       |A\setminus\tilde A|+|B|\leq\eta m,\quad
       |\tilde A|\geq|M_h(G,\xi)|-\theta n
    \right]\geq1-e^{-c_2n^{1/3}}.
    \label{eq:positive-conversion-performance}
\end{equation}
The coefficient seed is precisely $\xi$. In fact, each coordinate $f_e(G;\xi)$ is integer-valued whenever $G$ is a graph, so $B=\varnothing$ for every $G$ and $\xi$.
\end{proposition}

With these propositions in hand, we are ready to prove Theorem~\ref{thm:sparse-positive-low-degree}.

\begin{proof}[Proof of Theorem~\ref{thm:sparse-positive-low-degree}]
Use Proposition~\ref{prop:positive-local-precursor} to choose $Q$, let $r=2Q+4$, and let $h_e\coloneqq\bbone_{\{e\in I_Q\}}$, so that $|M_h(G,\xi)|=|I_Q|$.
Since the algorithm assigns colors to all potential edges independently of $G$, the color vector can serve as the coefficient seed $\xi$ in Proposition~\ref{prop:positive-edge-local-to-polynomial}; only the colors of present edges affect $h$.
We apply Proposition~\ref{prop:positive-edge-local-to-polynomial} with
\[
    \theta=\frac{\eps\log d}{128d},
    \qquad
    \eta=\frac{\theta n}{m}=\frac{\eps\Phi}{128m}.
\]
These choices fix the degree $D$ independently of $n$.
For a sufficiently small fixed $c_0>0$, outside an event of probability at most $e^{-c_0n^{1/3}}$, the conclusions of both propositions hold and the rounding test succeeds. Consequently, recalling the set $\mathcal{M}_f^\eta(G)$ from Definition~\ref{def:rounded-polynomial-output}, we have
\[
    |\mathcal M_f^\eta(G)|
    \geq |I_Q|-\theta n
    \geq\left(\frac12-\frac{\eps}{2}
                         -\frac{\eps}{128}\right)\Phi
    >\left(\frac12-\eps\right)\Phi.
\]
Set
\[
    \gamma\coloneqq\max\left\{1,\,
       \frac{\Gamma d}{(1/2-\eps)\log d}\right\}.
\]
By~\eqref{eq:positive-conversion-norm}, this choice ensures that $\E_G\|f(G;\xi)\|^2\leq\Gamma n\le\gamma(1/2-\eps)\Phi$ for every $\xi$.
Thus the seed-fixing argument of Lemma~\ref{lem:random_poly_to_deterministic} reduces to averaging the failure probability: since its average over the finite color seed is at most $e^{-c_0n^{1/3}}$, some seed $\xi_n$ satisfies the same bound. Fixing this seed proves the theorem with $c=c_0$.
\end{proof}

\subsection{Proof of Proposition~\ref{prop:positive-local-precursor}: a local algorithm for induced matchings}
\label{sec:positive-local-details}

We first establish a neighborhood estimate, which will also be used in the proof of Proposition~\ref{prop:positive-edge-local-to-polynomial}.
Write $Y_e=\bbone_{\{e\in E(G)\}}$ for the graph input coordinates.

\begin{lemma}
\label{lem:positive-edge-rooted-concentration}
Fix $d>1$ and an integer $R\geq1$. For every fixed integer $q\geq1$,
\begin{equation}
    \sup_{n>d,\,e}
    \E\bigl[(|\mathsf B_R(G,e)|+1)^q\mid Y_e=1\bigr]<\infty.
    \label{eq:positive-edge-ball-moments}
\end{equation}
Moreover, for every $\theta>0$, there exist a fixed integer $s\geq1$ and a constant $c>0$ such that, for all sufficiently large $n$,
\begin{equation}
    \Prb\left[
       |\{e\in E(G):|\mathsf B_R(G,e)|>s\}|>\theta n
    \right]\leq e^{-cn^{1/3}}.
    \label{eq:positive-edge-rooted-concentration}
\end{equation}
\end{lemma}

\begin{proof}
Conditional on $Y_e=1$, the breadth-first exploration of $G$ from the two endpoints of $e$, up to depth $R+1$, is stochastically dominated by a branching process with two initial individuals and independent $\mathrm{Bin}(n,d/n)$ offspring distributions.
These offspring distributions have uniformly bounded moments of every fixed order, and hence so does the total number $V_R$ of individuals in the first $R+1$ generations.
Every edge of $\mathsf B_R(G,e)$ joins two vertices reached by this exploration, so $|\mathsf B_R(G,e)|\leq V_R^2$, which proves~\eqref{eq:positive-edge-ball-moments}.

Let $Z_s\coloneqq|\{e\in E(G):|\mathsf B_R(G,e)|>s\}|$. Applying Markov's inequality to $|\mathsf B_R(G,e)|$ conditional on $Y_e=1$ for each potential edge $e$, and using~\eqref{eq:positive-edge-ball-moments} with $q=1$, we obtain $\E [Z_s]\leq C_{d,R}\,dn/(2s)$ for a constant $C_{d,R}$. Choose $s$ so that this is at most $\theta n/4$.

To show that $Z_s$ concentrates, we express it through a bounded local statistic. For a vertex $v$, let $g_s(G,v)$ denote the number of edges $e$ incident to $v$ with $|\mathsf B_R(G,e)|\le s$, and observe that
\[
    S_s\coloneqq\frac12\sum_v g_s(G,v)=|E(G)|-Z_s.
\]
The quantity $g_s(G,v)$ is determined by the ball of radius $R+1$ about $v$ in $G$. Moreover, $g_s\leq s$: if some edge $e$ incident to $v$ satisfies $|\mathsf B_R(G,e)|\le s$, then, since $R\geq 1$, the ball $\mathsf B_R(G,e)$ contains every edge incident to $v$, so $v$ has degree at most $s$. %

It therefore suffices to show that both $\sum_v g_s(G,v)$ and $|E(G)|$ concentrate. We apply~\cite[Corollary~3.5]{Wein2021} to the sum of the bounded local functionals $g_s(G,v)$, and a standard Chernoff bound to $|E(G)|$. Except with probability $e^{-cn^{1/3}}$, both quantities differ from their means by at most $\theta n/4$. Since $\E[Z_s]\leq \theta n/4$, it follows that $Z_s\leq\E[Z_s]+\theta n/2<\theta n$.
\end{proof}

Recall that the set $C$ in Algorithm~\ref{alg:finite-color-conflict} consists of the edges of $G$ that have a neighbor of the same color in $H=\mathrm{Line}(G)^2$.

\begin{lemma}
\label{lem:positive-monochromatic-conflicts}
For fixed $d>1$ and $\rho>0$, there is an integer $Q_0=Q_0(d,\rho)\geq2$ such that, for every $Q\geq Q_0$, some $c=c(d,\rho)>0$, and all sufficiently large $n$,
\[
    \Prb_{G,\chi}\left[|C|>\rho\Phi/2\right]\leq e^{-cn^{1/3}}.
\]
\end{lemma}

\begin{proof}
Apply Lemma~\ref{lem:positive-edge-rooted-concentration} with $R=4$ and $\theta=\rho\log d/(4d)$ to obtain a cutoff $s\geq 1$ such that, except with probability $e^{-cn^{1/3}}$, the set
\[
    U\coloneqq\{e\in E(G):|\mathsf B_4(G,e)|\leq s\}
\]
satisfies $|E(G)\setminus U|\leq\theta n=\rho\Phi/4$. Moreover, since $\E[|E(G)|]=(1-1/n)dn/2$, the Chernoff bound gives $|E(G)|\leq dn$ except with probability $e^{-\Omega_d(n)}$. Let
\[
    \mathcal{H}\coloneqq\{|E(G)|\leq dn\}\cap \{|E(G)\setminus U|\leq \rho \Phi/4\},
\]
and condition on a graph $G$ in $\mathcal H$.

Let $X\coloneqq|C\cap U|$. Each $e\in U$ has at most $s$ neighbors in $H$, since $N_H[e]=\mathsf B_2(G,e)\subseteq\mathsf B_4(G,e)$, and each of them shares the color of $e$ with probability $1/Q$. Hence
\[
    \E[X\mid G]\leq\frac{dns}{Q}\leq\frac{\rho\Phi}{8}
    \quad\text{whenever}\quad Q\geq Q_0\coloneqq\left\lceil\frac{8d^2s}{\rho\log d}\right\rceil.
\]

Conditional on $G$, the quantity $X$ is a function of the independent colors $(\chi_a)_{a\in E(G)}$; colors of absent edges do not affect it. Changing the color $\chi_a$ of a single edge $a\in E(G)$ can change the membership in $C$ only of edges in $N_H[a]$, and hence changes $X$ only through edges in $U\cap N_H[a]$. If this set is nonempty, then for any $e$ in it,
\[
    N_H[a]=\mathsf B_2(G,a)\subseteq\mathsf B_4(G,e),
\]
and $|\mathsf B_4(G,e)|\leq s$ since $e\in U$. Thus changing a single color changes $X$ by at most $s$. %
McDiarmid's inequality~\cite{mcdiarmid1989method} over the at most $dn$ independent colors therefore gives
\[
    \Prb\left[X>\rho\Phi/4\mid G\right]
    \leq\exp\left(-\frac{2(\rho\Phi/8)^2}{dns^2}\right)
    =e^{-\Omega_{d,\rho}(n)}.
\]
On $\mathcal H$, outside this event we have $|C|\leq X+|E(G)\setminus U|\leq\rho\Phi/4+\rho\Phi/4=\rho\Phi/2$. Since $\Prb[\mathcal H^c]\le e^{-cn^{1/3}}$ for a suitable $c>0$, the lemma follows.
\end{proof}

We are now ready to prove Proposition~\ref{prop:positive-local-precursor}.
We first show that, for every $Q\geq2$, the output is an induced matching determined by a bounded-radius ball, and that it is close to a maximal one. We then choose $Q$.

We begin by showing that the algorithm is local. The coloring $\chi$ is proper on $H[V(H)\setminus C]$: if two adjacent vertices of $H$ share a color, both lie in $C$. In round $j$, the algorithm accepts only vertices of color $j$ outside $C$, which form an independent set in $H$, and only those with no neighbor accepted in an earlier round. Hence $I_Q$ is an independent set in $H$, that is, an induced matching in~$G$.

To decide whether $e\in I_Q$, it suffices to know whether $e\in C$ and whether some neighbor $f$ of $e$ in $H$ with $\chi(f)<\chi(e)$ lies in $I_Q$. For example, if $\chi(e)=3$ and $e\notin C$, then $e\notin I_Q$ precisely when $e$ has a color-$1$ neighbor in $H\setminus C$, or a color-$2$ neighbor in $H\setminus C$ that has no color-$1$ neighbor in $I_1$. Unfolding this recursion, each step strictly decreases the color, so the resulting chains have length at most $Q-1$, and checking membership in $C$ along a chain requires one further step. Hence whether $e\in I_Q$ is determined by the colored ball of radius $Q+1$ about $e$ in $H$. %
Each step in $H$ corresponds to at most two steps in $\mathrm{Line}(G)$, and allowing two further steps to locate the edges witnessing conflict adjacencies, we conclude that whether $e\in I_Q$ is determined by the colored ball $\mathsf B_{2Q+4}(G,e)$.

Next, we show that the output is \emph{almost} maximal. Every vertex of $V(H)\setminus C$ not in $I_Q$ was rejected because it had a neighbor selected in an earlier round. Hence $I_Q$ is a maximal independent set in $H[V(H)\setminus C]$.
Extend $I_Q$ to a maximal independent set $J$ of $H$. Since every vertex of $V(H)\setminus (C\cup I_Q)$ already has a neighbor in $I_Q$, every added vertex lies in $C$, so
\[
    J\setminus I_Q\subseteq C \implies |I_Q|\geq|J|-|C|.
\]
In other words, although $I_Q$ need not be maximal, it is within $|C|$ of a maximal induced matching of $G$.

It remains to choose $Q$ so that $|J|$ is large and $|C|$ is small. Since $p=d/n$, we have $L=\Phi+O_d(1)$. For $d$ sufficiently large in terms of $\eps$, the proof of Proposition~\ref{prop:maximal-matching-half-optimal}, applied with $\eps/8$ in place of $\eps$ and with $\kappa=1/2$, shows that every maximal induced matching $J$ satisfies
\[
    |J|\geq\left(\frac12-\frac{\eps}{8}\right)L\geq\left(\frac12-\frac{\eps}{4}\right)\Phi
\]
for all sufficiently large $n$, except with probability at most $L\exp\bigl(-(np)^{\eps/4}/(30p)\bigr)=e^{-\Omega_{d,\eps}(n)}$. %
Next, Lemma~\ref{lem:positive-monochromatic-conflicts} with $\rho=\eps/2$ gives $Q_0=Q_0(d,\eps/2)$ such that, for $Q\geq Q_0$, we have $|C|\leq\eps\Phi/4$ except with probability $e^{-cn^{1/3}}$. Fixing such a $Q$, we conclude that, except with probability $e^{-c_1n^{1/3}}$ for a suitable $c_1>0$,
\[
    |I_Q|\geq
    \left(\frac12-\frac{\eps}{4}\right)\Phi
       -\frac{\eps}{4}\Phi
    =\left(\frac12-\frac{\eps}{2}\right)\Phi,
\]
completing the proof.

\subsection{Proof of Proposition~\ref{prop:positive-edge-local-to-polynomial}: from local to low-degree}
\label{sec:positive-conversion-details} 

Let $s$ be the cutoff given by Lemma~\ref{lem:positive-edge-rooted-concentration} with $R=r+2$ and the given $\theta$.

We first construct a polynomial that reproduces $h$ exactly at edges with small neighborhoods. A rule $h$ of bounded radius does not by itself give a polynomial of bounded degree, since the ball $\mathsf B_r(G,e)$ may contain many edges. %
Instead, we use the rule to define the coefficients of a polynomial of bounded degree that agrees with $h$ whenever the relevant ball is small. %
For a potential edge $e$, let $\mathscr F_{e,r,s}$ be the family of sets $F\subseteq\binom{[n]}{2}$ of potential edges satisfying
\[
    e\in F,\qquad |F|\leq s,\qquad
    \mathrm{dist}_{\mathrm{Line}(F)}(e,a)\leq r
       \quad\text{for every }a\in F,
\]
where $\mathrm{dist}_{\mathrm{Line}(F)}$ denotes the graph distance in the line graph of the graph $([n],F)$.
For $F \in \mathscr F_{e,r,s}$, we evaluate $h_e(F,\xi)$ on the graph $([n],F)$, whose ball of radius $r$ about $e$ is $F$ itself, and define recursively %
\begin{equation}
    \alpha_e(F;\xi)\coloneqq h_e(F,\xi)-
       \sum_{\substack{F'\in\mathscr F_{e,r,s}\\F'\subsetneq F}}
                \alpha_e(F';\xi).
    \label{eq:positive-mobius-coefficients}
\end{equation}
That is, we first set $\alpha_e(\{e\};\xi)=h_e(\{e\},\xi)$, then define $\alpha_e(F;\xi)$ for the sets $F=\{e,a\}$ with $a$ sharing an endpoint with $e$, and so on in order of increasing $|F|$.

For $Y=(Y_a)_{a\in\binom{[n]}{2}}\in \RR^m$, set
\begin{equation}
    f_e(Y;\xi)\coloneqq \sum_{F\in\mathscr F_{e,r,s}}
          \alpha_e(F;\xi)\prod_{a\in F}Y_a.
    \label{eq:positive-edge-root-polynomial}
\end{equation}
Each $f_e$ has degree at most $s$. Every monomial contains $Y_e$, and since $h$ is $\{0,1\}$-valued, the recursion~\eqref{eq:positive-mobius-coefficients} produces integer coefficients. Hence, on graph inputs $Y\in\{0,1\}^m$,
\begin{equation}
    Y_e=0\ \Longrightarrow\ f_e(Y;\xi)=0,
    \qquad f_e(Y;\xi)\in\mathbb Z.
    \label{eq:positive-root-factor}
\end{equation}
In particular, absent edges $e\notin E(G)$ never belong to $A$, and since every score is an integer, the uncertain set $B$ is always empty.

On a graph input $G$, the monomial indexed by $F$ survives precisely when $F\subseteq E(G)$, in which case every path from $e$ in $\mathrm{Line}(F)$ is also a path in $\mathrm{Line}(G)$, so $F\subseteq\mathsf B_r(G,e)$. Now suppose that $e\in E(G)$ and $|\mathsf B_r(G,e)|\leq s$. Then $\mathsf B_r(G,e)$ itself belongs to $\mathscr F_{e,r,s}$, since shortest paths from $e$ to edges of the ball stay inside the ball. Applying~\eqref{eq:positive-mobius-coefficients} with $F=\mathsf B_r(G,e)$, we obtain
\begin{equation} \label{eq:positive-polynomial-exactness}
    f_e(G;\xi)
    =\sum_{\substack{F\in\mathscr F_{e,r,s}\\F\subseteq\mathsf B_r(G,e)}}
         \alpha_e(F;\xi)
    =\alpha_e(\mathsf B_r(G,e);\xi)+\sum_{\substack{F\in\mathscr F_{e,r,s}\\F\subsetneq \mathsf B_r(G,e)}}\alpha_e(F;\xi)
    =h_e(\mathsf B_r(G,e),\xi)=  h_e(G,\xi).
\end{equation}

Next, we show that most selected edges survive rounding.
Let $Z$ be the set of edges $e\in E(G)$ for which $\mathsf B_{r+2}(G,e)$ has more than $s$ edges.
If $e\in E(G)\setminus Z$, then~\eqref{eq:positive-polynomial-exactness} holds at $e$ and at every neighbor of $e$ in $H(G)$, because
\[
    a\sim_{H(G)}e
       \quad\Longrightarrow\quad
    \mathsf B_r(G,a)\subseteq\mathsf B_{r+2}(G,e).
\]
Let $e\in E(G)\setminus Z$. If $e\in A$, then $h_e(G,\xi)=f_e(G;\xi)\ge 1$ by~\eqref{eq:positive-polynomial-exactness}, so $e\in M_h(G,\xi)$. Conversely, if $e\in M_h(G,\xi)$, then $f_e(G;\xi)=h_e(G,\xi)=1$, so $e\in A$. In either case, no neighbor $a$ of $e$ in $H(G)$ lies in $A$: otherwise, since~\eqref{eq:positive-polynomial-exactness} also holds at $a$, we would have $h_a(G,\xi)=1$, contradicting the assumption that $M_h(G,\xi)$ is an induced matching. %
Hence every $e\in (A\cup M_h(G,\xi))\setminus Z$ lies in $\tilde A$, and
\begin{equation}
    A\setminus\tilde A\subseteq Z,\qquad
    B=\varnothing,\qquad M_h(G,\xi)\setminus Z\subseteq\tilde A.
    \label{eq:positive-rounding-safe}
\end{equation}
By Lemma~\ref{lem:positive-edge-rooted-concentration} with $R=r+2$, we have $|Z|\leq\theta n=\eta m$ except with probability $e^{-cn^{1/3}}$, which proves~\eqref{eq:positive-conversion-performance}.

Finally, we bound the norm of $f$ uniformly in the seed. By induction on $|F|$, the recursion~\eqref{eq:positive-mobius-coefficients} gives $|\alpha_e(F;\xi)|\leq a_s$ for a constant $a_s$ depending only on $s$, since each $F$ has at most $2^s$ subsets.
Let $N_e\coloneqq|\mathsf B_r(G,e)|$ if $Y_e=1$, and $N_e\coloneqq0$ otherwise.
Since every surviving monomial corresponds to a subset of $\mathsf B_r(G,e)$ of size at most $s$, there are at most $(N_e+1)^s$ of them, so
\begin{equation}
    |f_e(G;\xi)|^2\leq Y_e\,a_s^2(N_e+1)^{2s}.
    \label{eq:positive-coordinate-moment-bound}
\end{equation}
The moment bound~\eqref{eq:positive-edge-ball-moments}, with $R=r$ and $q=2s$, gives, uniformly in the seed,
\[
    \E_G f_e(G;\xi)^2\leq C_{d,r,s}\frac dn,
    \qquad
    \E_G\|f(G;\xi)\|^2
       \leq\binom n2 C_{d,r,s}\frac dn\leq\Gamma n.
\]
The factor $d/n$ comes from the variable $Y_e$ present in every monomial; it keeps the total squared norm of the $\Theta(n^2)$ coordinates of order $n$. This proves~\eqref{eq:positive-conversion-norm}.

\section{Low-degree hardness: proof of Theorem~\ref{thm:sparse_low_degree_hardness}}
\label{sec:low-degree}

In this section, we prove Theorem~\ref{thm:sparse_low_degree_hardness}.
Throughout, $d$ is fixed and $n\to\infty$.

\subsection{Forbidden structures}

We now state and prove the forbidden structure result outlined in Section~\ref{subsec:proof-techniques}.
We first give a high-probability upper bound on the maximum induced matching size, based on~\cite{CooleyDraganicKangSudakov2021}, which allows us to restrict attention to induced matchings with at most $(1+\eps)\Phi$ edges.
\begin{lemma}
\label{lem:sparse_largest_IM}
    Fix $\eps > 0$.
    If the fixed constant $d = d(\eps)$ is sufficiently large, then there is $c=c(d,\eps)>0$ such that, for all sufficiently large $n$,
    \[
        \Prb\left[\text{$G(n,d/n)$ has an induced matching with more than $(1+\eps)\Phi$ edges}\right]\leq e^{-cn}.
    \]
\end{lemma}

\begin{proof}
For an integer $r\geq1$, let $Y_r$ denote the number of induced matchings with exactly $r$ edges in $G(n,p)$.
By \cite[Section~2, Claim~1, p.~269]{CooleyDraganicKangSudakov2021},
\begin{equation}
    \E[Y_r]
    =\binom n{2r}\frac{(2r)!}{2^r r!}
      p^r(1-p)^{\binom{2r}{2}-r}.
    \label{eq:sparse-largest-count}
\end{equation}
Indeed, the factors choose and pair the $2r$ endpoints, require the $r$ matching edges to be present, and require every other pair of endpoints to be absent.

Put $p=d/n$ and $r=\lfloor(1+\eps)\Phi\rfloor+1$.
If $r>n/2$, the event in the lemma is empty, so we may assume $r\leq n/2$.
Applying~\eqref{eq:sparse-largest-count} together with $\binom n{2r}(2r)!\leq n^{2r}$, $r!\geq(r/e)^r$, and $1-x\leq e^{-x}$ gives
\begin{equation}
    \log\E[Y_r]
    \leq r\left(
       2\log n+\log(d/n)+1-\log(2r)
       -2\frac dn(r-1)
    \right).
    \label{eq:sparse-largest-first-moment}
\end{equation}
Writing $a=r/\Phi>1+\eps$, the expression in parentheses equals
\[
    2(1-a)\log d-\log(2a\log d)+1+\frac{2d}{n}.
\]
This expression is decreasing in $a$, and, taking first $d$ and then $n$ sufficiently large, it is at most $-\eps\log d$ uniformly for $a\geq1+\eps$.
Any induced matching with more than $(1+\eps)\Phi$ edges contains an induced matching with exactly $r$ edges.
Thus Markov's inequality gives
\[
    \Prb\left[\text{$G(n,d/n)$ has an induced matching with more than $(1+\eps)\Phi$ edges}\right]
    \leq \E[Y_r]
    \leq e^{-\eps r\log d}
    \leq e^{-cn},
\]
where $c=\eps(1+\eps)(\log d)^2/d>0$.
\end{proof}

We now define the sequence of correlated random graphs used to construct the forbidden structure.
\begin{definition}\label{def:sparse_interpolation_path}
    Given $n \in \NN$, fix an arbitrary ordering $e_1,\dots,e_m$ of the set $\mathcal E_n\coloneqq\binom{[n]}{2}$ of potential edges.
    Choose $G^{(0)}\sim G(n,d/n)$ and mutually independent random variables $(Z_t)_{t\geq1}$ with law $\mathrm{Bernoulli}(d/n)$, independent of $G^{(0)}$.
    For $t\geq1$, obtain $G^{(t)}$ from $G^{(t-1)}$ by replacing the indicator of the potential edge
    \[
        e_{\sigma(t)},
        \qquad\text{where}\qquad
        \sigma(t)=1+((t-1)\bmod m),
    \]
    by $Z_t$, and retaining every other coordinate.
    We call the graphs $G^{(0)},\ldots,G^{(T)}$ a \emph{length-$T$ interpolation path}.
\end{definition}

Note that every $G^{(t)}$ has law $G(n,d/n)$.
Moreover, in any $m=\binom n2$ consecutive steps, every edge indicator is resampled exactly once, so $G^{(t+m)}$ is independent of $G^{(0)},\ldots,G^{(t)}$ for every $t\geq0$.

The next result gives the forbidden structure that drives the hardness proof.
\begin{proposition}\label{prop:sparse_endpoint_forbidden_structure}
    Fix $0<\eps\leq1/2$, an exponent $A<\infty$, and an integer $K\geq 4/\eps^2$.
    Consider an interpolation path $G^{(0)},\dots,G^{(T)}$ as in Definition~\ref{def:sparse_interpolation_path}, where $0\leq T\leq n^A$.
    If the fixed constant $d=d(\eps,K,A)$ is sufficiently large, then there is $c=c(\eps,K,A,d)>0$ such that, for all sufficiently large $n$, with probability at least $1-e^{-cn}$, there do not exist $S_1,\dots,S_K\subseteq\mathcal E_n$ satisfying the following properties:
    \begin{enumerate}[label=(\roman*)]
        \item\label{item:forbidden-time} for each $k \in [K]$, there exists $0 \leq t_k \leq T$ such that $S_k$ is an induced matching in $G^{(t_k)}$;
        \item\label{item:forbidden-size} $\abs{S_k} \geq \inparen{\frac{1}{2}+\eps}\Phi$ for all $k \in [K]$;
        \item\label{item:forbidden-new} $\abs{V(S_k) \setminus \bigcup_{\ell < k} V(S_\ell)} \in \insquare{\frac{\eps}{4},\frac{\eps}{2}}\Phi$ for all $2 \leq k \leq K$.
    \end{enumerate}
\end{proposition}

\begin{proof}
Throughout this proof, implicit constants in $O(\cdot)$ may depend on $\eps$ and $K$, and the constants in $\exp(-\Omega(n))$ may also depend on $d$.

By Lemma~\ref{lem:sparse_largest_IM} and a union bound over the interpolation path, with probability $1-\exp(-\Omega(n))$ no $G^{(t)}$ contains an induced matching with more than $(1+\eps)\Phi$ edges.
It therefore suffices to rule out forbidden structures that additionally satisfy
\begin{enumerate}[label=(\roman*), start=4]
    \item\label{item:forbidden-upper} $\abs{S_k}\leq (1+\eps)\Phi$ for all $k\in[K]$.
\end{enumerate}

We use a first moment argument.
Let $X$ be the number of tuples $(t_1,S_1),\dots,(t_K,S_K)$ satisfying \ref{item:forbidden-time}--\ref{item:forbidden-upper}.
We show that $\E[X]=\exp(-\Omega(n))$, after which Markov's inequality completes the proof.

For $2 \leq k \leq K$, define $\Delta_k \coloneqq V(S_k) \setminus \bigcup_{\ell < k} V(S_\ell)$.
Write $\abs{S_k} = a_k \Phi$ and $\abs{\Delta_k} = b_k \Phi$.
Properties \ref{item:forbidden-size}, \ref{item:forbidden-new}, and \ref{item:forbidden-upper} restrict $a_k \in \insquare{\frac{1}{2} + \eps, 1 + \eps}$ and $b_k \in \insquare{\frac{\eps}{4}, \frac{\eps}{2}}$, and we always assume these constraints below.
Write
\[
    \ab=(a_1,\ldots,a_K),\qquad
    \bbb=(b_2,\ldots,b_K),\qquad\text{and}\qquad
    \tb=(t_1,\ldots,t_K),
\]
where the times $t_1,\ldots,t_K$ are as in \ref{item:forbidden-time} and need not be in increasing order.
Let $\mathcal F(\ab,\bbb)$ be the family of ordered tuples $\Sbb=(S_1,\ldots,S_K)$ of matchings on $[n]$ with the prescribed values of $\abs{S_k}$ and $\abs{\Delta_k}$.
Then
\begin{align}\label{eq:forbiddenStructure_EX_decomposition}
    \E[X]
    &=\sum_{\ab,\bbb,\tb}
      \sum_{\Sbb\in\mathcal F(\ab,\bbb)}
      \Prb\left[S_k\text{ is an induced matching in }G^{(t_k)}\text{ for all }k\in[K]\right]
      \nonumber\\
    &\leq n^{O(K)}(T+1)^K\sup_{\ab,\bbb,\tb}
      \underbrace{\sum_{\Sbb\in\mathcal F(\ab,\bbb)}
      \Prb\left[S_k\text{ is an induced matching in }G^{(t_k)}\text{ for all }k\in[K]\right]}_{\eqqcolon C(\ab,\bbb,\tb)},
\end{align}
where we have bounded the number of admissible sequences $\ab$ and $\bbb$ by $n^{O(K)}$.
It remains to bound $C(\ab,\bbb,\tb)$ uniformly over $\ab,\bbb,\tb$.

For a tuple $\Sbb$, let
\[
    A_k\coloneqq\{S_k\text{ is an induced matching in }G^{(t_k)}\},
    \qquad\text{and}\qquad H_j\coloneqq\bigcap_{\ell=1}^j A_\ell.
\]
We have the decomposition
\begin{align}\label{eq:forbiddenStructure_conditional_sums}
    C(\ab,\bbb,\tb)
    &=\sum_{S_1}\Prb(A_1)
      \sum_{S_2}\Prb(A_2\mid H_1)\cdots
      \sum_{S_K}\Prb(A_K\mid H_{K-1})
      \nonumber\\
    &\leq\left(\sum_{S_1}\Prb(A_1)\right)
      \prod_{k=2}^K\left[
      \sup_{\substack{S_1,\ldots,S_{k-1}\\\Prb(H_{k-1})>0}}
      \sum_{S_k}\Prb(A_k\mid H_{k-1})\right].
\end{align}
Here the sums in the first line are nested: the first runs over matchings $S_1$ of size $a_1\Phi$, and, for $k\geq2$, the $k$-th runs over matchings $S_k$ of size $a_k\Phi$ with exactly $b_k\Phi$ endpoints outside $\bigcup_{\ell<k}V(S_\ell)$.
In the second line, each supremum is over all admissible earlier matchings $S_1,\ldots,S_{k-1}$ with $\Prb(H_{k-1})>0$.
We first bound the sum over $S_1$, and then bound each conditional sum uniformly over all such earlier matchings.

Since $G^{(t_1)}\sim G(n,d/n)$, applying~\eqref{eq:sparse-largest-count} and~\eqref{eq:sparse-largest-first-moment} with $r=a_1\Phi$ gives
\begin{align}\label{eq:forbiddenStructure_S1_contrib}
    \sum_{S_1}\Prb(A_1) &= \sum_{\substack{S_1\text{ a matching on }[n]\\|S_1|=a_1\Phi}}
    \Prb\left[S_1\text{ is an induced matching in }G^{(t_1)}\right]
    \nonumber\\
    &\leq\exp\inparen{\inparen{2a_1-2a_1^2+o_d(1)+o(1)}\Phi\log d}.
\end{align}

We now bound the remaining factors in~\eqref{eq:forbiddenStructure_conditional_sums}.
Fix $k\geq2$ and admissible matchings $S_1,\dots,S_{k-1}$ with $\Prb(H_{k-1})>0$, and condition on $H_{k-1}$.
Let
\[
    V_{<k}\coloneqq\bigcup_{\ell<k}V(S_\ell)
\]
denote the set of \emph{old endpoints}, and note that $\abs{V_{<k}}\leq 2K(1+\eps)\Phi$.

We first choose the endpoints of $S_k$.
The $b_k\Phi$ new endpoints, those outside $V_{<k}$, can be chosen in at most
\[
   \binom{n-\abs{V_{<k}}}{b_k\Phi}\leq\binom n{b_k\Phi}
   \leq\exp\inparen{\inparen{b_k+o_d(1)}\Phi\log d}
\]
ways.
The remaining endpoints of $S_k$ lie in $V_{<k}$, and there are at most
\[
    \binom{\abs{V_{<k}}}{(2a_k-b_k)\Phi}
    \leq2^{\abs{V_{<k}}}=\exp(o_d(\Phi\log d))
\]
ways to choose them.
Once the endpoints are fixed, it remains to choose the edges of $S_k$, which are of two types: (1) edges with at least one endpoint in $\Delta_k$, and (2) edges with both endpoints in $V_{<k}$.
For an upper bound, we may also count choices that do not form a matching.

We first account for edges of type (1).
Let $P_k\coloneqq\inbraces{e\in\binom{V(S_k)}2:e\cap\Delta_k\neq\varnothing}$ be the set of all pairs of endpoints that meet $\Delta_k$.
Then
\begin{equation}\label{eq:forbiddenStructure_Pk_count}
    \abs{P_k}
    =b_k\Phi(2a_k\Phi-b_k\Phi)+\binom{b_k\Phi}2
    =\inparen{2a_kb_k-\frac{b_k^2}2}\Phi^2+O(\Phi).
\end{equation}
For distinct potential edges $e$, the sequences $(G_e^{(t)})_{t\geq0}$ are independent.
The event $H_{k-1}$ constrains specified entries of these sequences edge by edge, so they remain independent after conditioning.
Every pair in $P_k$ has an endpoint outside $V_{<k}$, so $H_{k-1}$ imposes no constraint on its sequence.
Thus the indicators $(G_e^{(t_k)})_{e\in P_k}$ are still independent $\mathrm{Bernoulli}(d/n)$ variables.

Let $r$ be the number of edges of $S_k$ that meet $\Delta_k$, so that $r\leq b_k\Phi$.
Summing over the choices of these edges, and requiring them to be present and all other pairs in $P_k$ to be absent, we claim that the resulting contribution is at most
\begin{equation}\label{eq:forbiddenStructure_new_edges}
    \sum_{r=0}^{b_k\Phi}\binom{|P_k|}r
       \left(\frac dn\right)^r
       \left(1-\frac dn\right)^{|P_k|-r}
    \leq\exp\inparen{\inparen{-2a_kb_k+\frac{b_k^2}2+o(1)+o_d(1)}\Phi\log d}.
\end{equation}
To justify this estimate, note that for $N=O(\Phi^2)$ and $1\leq r\leq R=O(\Phi)$,
\[
    \binom Nr\left(\frac dn\right)^r
    \leq\left(\frac{eNd}{rn}\right)^r.
\]
Using $d\Phi/n=\log d$ and $r\log(\Phi/r)\leq\Phi/e$, and summing over the $O(\Phi)$ possible values of $r$, we obtain
\begin{equation}\label{eq:forbiddenStructure_binomial_sum}
    \sum_{r=0}^{\lfloor R\rfloor}\binom Nr\left(\frac dn\right)^r
    \leq\exp\inparen{O_{\eps,K}(\Phi\log\log d+\log n)}.
\end{equation}
Moreover, uniformly for $r\leq b_k\Phi$, we have $(1-d/n)^{-r}=\exp(O(\log d))$, while~\eqref{eq:forbiddenStructure_Pk_count} gives
\[
    \left(1-\frac dn\right)^{|P_k|}
    =\exp\inparen{-\inparen{2a_kb_k-\frac{b_k^2}2+o(1)}\Phi\log d}.
\]
Together with~\eqref{eq:forbiddenStructure_binomial_sum}, this proves~\eqref{eq:forbiddenStructure_new_edges}.

It remains to account for edges of type (2), whose endpoints both lie in $V_{<k}$.
Let
\[
    Q_{<k}\coloneqq\bigcup_{\ell<k}S_\ell,
    \qquad\text{so that}\qquad |Q_{<k}|\leq K(1+\eps)\Phi.
\]
For $e\notin Q_{<k}$, the constraints in $H_{k-1}$ can only require $e$ to be absent.
Hence the indicator $G_e^{(t_k)}$ is either forced to be zero by $H_{k-1}$ or is a fresh $\mathrm{Bernoulli}(d/n)$ variable independent of $H_{k-1}$; in either case, its conditional probability of being present is at most $d/n$.
For $e\in Q_{<k}$, we use the trivial upper bound $1$.
Since, for any $F\subseteq\binom{V_{<k}}2$, the indicators $(G_e^{(t_k)})_{e\in F}$ are conditionally independent given $H_{k-1}$,
\begin{align}\label{eq:forbiddenStructure_old_presence}
    \Prb\left[F\subseteq E(G^{(t_k)})\mid H_{k-1}\right]
    &=\prod_{e\in F}\Prb\left[G_e^{(t_k)}=1\mid H_{k-1}\right]
    \nonumber\\
    &\leq\left(\frac dn\right)^{|F\setminus Q_{<k}|}.
\end{align}
There are at most $2^{|Q_{<k}|}$ ways to choose the edges of $S_k$ that lie in $Q_{<k}$.
Every other edge of $S_k$ of type (2) has conditional presence probability at most $d/n$, and there are at most $|S_k|\leq(1+\eps)\Phi$ such edges.
Ignoring the remaining absence constraints and applying~\eqref{eq:forbiddenStructure_binomial_sum}, the total contribution of edges of type (2) is at most
\begin{equation}\label{eq:forbiddenStructure_old_edges}
    2^{|Q_{<k}|}
    \sum_{r=0}^{\lfloor(1+\eps)\Phi\rfloor}
    \binom{\binom{|V_{<k}|}2}r\left(\frac dn\right)^r
    \leq\exp\inparen{\inparen{o_d(1)+o(1)}\Phi\log d}.
\end{equation}

Altogether, summing over matchings $S_k$ with $|S_k|=a_k\Phi$ and $|V(S_k)\setminus V_{<k}|=b_k\Phi$, we obtain
\begin{align}\label{eq:forbiddenStructure_step}
    \sum_{S_k}\Prb(A_k\mid H_{k-1}) &=\sum_{S_k}\Prb\left[S_k\text{ is an induced matching in }G^{(t_k)}\mid H_{k-1}\right]
    \nonumber\\
    &\leq\exp\inparen{\inparen{b_k-2a_kb_k+\frac{b_k^2}2+o(1)+o_d(1)}\Phi\log d}.
\end{align}
This bound is uniform over all admissible earlier matchings $S_1,\ldots,S_{k-1}$, and all $o(1)$ and $o_d(1)$ terms above are uniform over the admissible $a_k,b_k$ and over $t_1,\ldots,t_K\in\{0,\ldots,T\}$.

Applying~\eqref{eq:forbiddenStructure_S1_contrib} and~\eqref{eq:forbiddenStructure_step} to~\eqref{eq:forbiddenStructure_conditional_sums} bounds $C(\ab,\bbb,\tb)$, and substituting into~\eqref{eq:forbiddenStructure_EX_decomposition} gives
\begin{align}\label{eq:forbiddenStructure_total}
    \E[X]
    &\leq
    n^{O(K)}(T+1)^K
    \nonumber\\
    &\qquad\times
    \sup_{\ab,\bbb}
    \exp\inparen{
        \inparen{
            2a_1-2a_1^2
            +\sum_{k=2}^K
            \inparen{
                b_k-2a_kb_k+\frac{b_k^2}{2}
            }
            +o(1)+o_d(1)
        }\Phi\log d
    }.
\end{align}
We have $2a_1-2a_1^2\leq 1/2$.
Moreover, since $a_k\geq 1/2+\eps$ and $\eps/4\leq b_k\leq\eps/2$,
\[
    b_k-2a_kb_k+\frac{b_k^2}{2}
    \leq
    -2\eps b_k+\frac{b_k^2}{2}
    \leq
    -\frac{7}{4}\eps b_k
    \leq
    -\frac{7}{16}\eps^2.
\]
Summing over the $K-1$ values $k=2,\ldots,K$ gives
\[
    2a_1-2a_1^2
    +\sum_{k=2}^K\left(b_k-2a_kb_k+\frac{b_k^2}{2}\right)
    \leq\frac12-\frac7{16}(K-1)\eps^2.
\]
Since $0<\eps\leq1/2$ and $K\geq 4/\eps^2$, we have $(K-1)\eps^2\geq 4-\eps^2\geq 15/4$, so the right-hand side is at most $1/2-105/64<-1$.
Thus the coefficient of $\Phi\log d$ in~\eqref{eq:forbiddenStructure_total}, excluding the $o(1)+o_d(1)$ terms, is less than $-1$.
Taking $d$ sufficiently large to absorb the uniform $o_d(1)$ terms, and then $n$ sufficiently large to absorb the $o(1)$ terms and the factor $n^{O(K)}(T+1)^K\leq n^{O(K)+AK}$, we obtain $\E[X]=\exp(-\Omega(n))$, since $\Phi\log d=(\log^2d/d)n=\Theta_d(n)$.
This completes the proof.
\end{proof}

\subsection{Stability of low-degree polynomials}

In this subsection, we give a stability result for low-degree polynomials along the interpolation path from Definition~\ref{def:sparse_interpolation_path}.
It states that, with sufficiently large probability, no single step of the interpolation path causes a large jump in the output of a low-degree polynomial.

Throughout this subsection, we identify a graph $G$ on $[n]$ with its edge indicator vector in $\{0,1\}^m$, with coordinates ordered as in Definition~\ref{def:sparse_interpolation_path}.
We begin by defining what it means for a single resampling step to create a large jump.

\begin{definition}\label{def:sparse_bad_update}
    Let $f:\RR^m\to\RR^m$ be a deterministic polynomial of degree at most $D$, let $\theta>0$, and put
    \[
        \mathcal E_f\coloneqq \E_{H\sim G(n,d/n)}\norm{f(H)}^2.
    \]
    A pair of graphs $(G,G')$ on $[n]$ is \emph{$\theta$-bad} for $f$ if $G$ and $G'$ differ in at most one potential edge and
    \[
        \norm{f(G)-f(G')}^2
        \geq
        \theta\mathcal E_f.
    \]
    An interpolation step $G^{(t-1)}\to G^{(t)}$ is \emph{$\theta$-bad} if the pair $(G^{(t-1)},G^{(t)})$ is $\theta$-bad.
\end{definition}

\begin{proposition}[Essentially {\cite[Proposition~2.6]{Wein2021}}]\label{prop:sparse_low_degree_stability}
    Let $\ell\in\NN$ and $\theta>0$, and suppose $d/n\leq 1/2$.
    Let $G^{(0)},\dots,G^{(T)}$ be an interpolation path as in Definition~\ref{def:sparse_interpolation_path} with $T=\ell m$.
    If $f:\RR^m\to\RR^m$ is a deterministic polynomial of degree at most $D$ with $\mathcal E_f>0$, then
    \[
        \Prb\left[
            \text{no interpolation step }G^{(t-1)}\to G^{(t)}
            \text{ is }\theta\text{-bad for }f
        \right]
        \geq
        \inparen{\frac dn}^{4\ell D/\theta}.
    \]
\end{proposition}

Proposition~\ref{prop:sparse_low_degree_stability} follows from~\cite[Proposition~2.6]{Wein2021} applied with $p=d/n$, $L=\ell$, and $c=\theta$.
Although Wein states the result for $n$ output coordinates, the same proof applies to $m$ output coordinates, since the estimates in \cite[Lemmas~2.7--2.8, equations~(6)--(7)]{Wein2021} hold for any number of output coordinates, with the same constants.

\subsection{Endpoints outside a fixed set}

For any fixed set $W$ of size $O(\Phi)$, with high probability every large induced matching has many endpoints outside $W$.
In the hardness proof, $W$ will consist of the endpoints of previously selected matchings, and we will apply this bound after all potential edges have been resampled along the interpolation path.

\begin{lemma}\label{lem:sparse_endpoint_avoidance_checkpoint}
    Fix $\eps>0$, $0\leq B<\infty$, and $0 < \mu < 2\inparen{\frac12+\eps}^2$.
    Let $W\subseteq[n]$ be a fixed set with $\abs{W}\leq B\Phi$, and let $G\sim G(n,d/n)$.
    If the fixed constant $d=d(\eps,B,\mu)$ is sufficiently large, then there is $c=c(\eps,B,\mu,d)>0$ such that, for all sufficiently large $n$,
    \[
        \Prb\left[
            \exists\text{ induced matching }M\text{ in }G:
            \inparen{\frac12+\eps}\Phi\leq \abs{M}\leq (1+\eps)\Phi,\,
            \abs{V(M)\setminus W}\leq \mu\Phi
        \right]\leq
        e^{-cn},
    \] 
    uniformly over all such sets $W$.

    Consequently, if $\cF$ is a sigma-field independent of $G$ and $W$ is $\cF$-measurable with $|W|\leq B\Phi$ almost surely, then the same bound holds conditionally on $\cF$, almost surely.
\end{lemma}

\begin{proof}
    We use a first moment argument.
    Let $X$ be the number of induced matchings satisfying the conditions in the event above.
    We show that $\E[X]=\exp(-\Omega(n))$, after which Markov's inequality completes the proof.

    Write $\abs{M}=a\Phi$ and $\abs{V(M)\setminus W}=b\Phi$, where $a$ and $b$ range over
    \[
        \frac12+\eps\leq a\leq 1+\eps,
        \qquad\text{and}\qquad
        0\leq b\leq \min\{\mu,2a\},
    \]
    with $a\Phi$ and $b\Phi$ integers.
    Then
    \[
        \E[X]
        \leq
        \sum_{a,b}
        \binom{n}{b\Phi}
        \binom{|W|}{(2a-b)\Phi}
        (2a\Phi)^{a\Phi}
        \inparen{\frac dn}^{a\Phi}
        \inparen{1-\frac dn}^{\binom{2a\Phi}{2}-a\Phi},
    \]
    where the first two factors choose the endpoints, the third bounds the number of pairings, and the last two require exactly the matching edges to be present among these endpoints.

    To bound the sum, we use
    \[
        \binom{|W|}{(2a-b)\Phi}\leq 2^{B\Phi},
        \qquad\text{and}\qquad
        (2a\Phi)^{a\Phi}\inparen{\frac dn}^{a\Phi}
        =(2a\log d)^{a\Phi},
    \]
    so the logarithms of these factors are $o_d(\Phi\log d)$.
    The factor $\binom{n}{b\Phi}$ has logarithm at most $(b+o_d(1))\Phi\log d$, while the probability that all other pairs of endpoints are absent has logarithm $-(2a^2+o(1))\Phi\log d$.
    Combining these estimates gives
    \begin{align*}
        \E[X]
        &\leq
        \sum_{a,b}
        \exp\inparen{
            \inparen{b-2a^2+o_d(1)+o(1)}\Phi\log d
        }
        \\
        &\leq
        n^2
        \exp\inparen{
            \inparen{
                \mu
                -2\inparen{\frac12+\eps}^2
                +o_d(1)+o(1)
            }\Phi\log d
        }.
    \end{align*}
    Since $\mu<2(1/2+\eps)^2$, taking $d$ sufficiently large and then $n$ sufficiently large makes the coefficient of $\Phi\log d$ negative.
    Since $\Phi\log d=\Theta_d(n)$, this gives the desired bound on $\E [X]$.
    Every estimate used $W$ only through the bound $|W|\leq B\Phi$, which proves the uniform and conditional statements.
\end{proof}

\subsection{Seed fixing}

The following lemma is used in both the hardness and achievability arguments.
It reduces a polynomial with a random coefficient seed to a deterministic polynomial, while controlling its normalization and failure probability.

\begin{lemma}[Essentially {\cite[Lemma~2.11]{Wein2021}}]\label{lem:random_poly_to_deterministic}
    Suppose $f(\,\cdot\,;\xi)$ is a random polynomial of degree at most $D$ whose coefficient seed $\xi$ is independent of $G$, and suppose it is a $(k,\delta,\gamma,\eta)$-optimizer for the induced matching problem in $G(n,d/n)$.
    Then for any $\lambda > 2$, there exists a deterministic seed $\xi_0$ for which $f(\,\cdot\,;\xi_0)$ is a deterministic
    \[
        \bigl(k,\min\{\lambda\delta,1\},\lambda\gamma,\eta\bigr)\text{-optimizer}
    \]
    of degree at most $D$.
\end{lemma}

\begin{proof}
Set
\[
    X(\xi)\coloneqq \frac{\E_G\|f(G;\xi)\|^2}{\gamma k},
    \qquad\text{and}\qquad
    q_{\mathrm{fail}}(\xi)\coloneqq \Prb_G\left[|\mathcal M_{f(\,\cdot\,;\xi)}^\eta(G)|<k\right].
\]
By Tonelli's theorem and the optimizer assumptions, $\E_\xi X(\xi)\leq1$ and $\E_\xi q_{\mathrm{fail}}(\xi)\leq\delta$.
If $\delta>0$, put $Y(\xi)\coloneqq q_{\mathrm{fail}}(\xi)/\delta$.
Markov's inequality and a union bound give
\[
    \Prb_\xi\left[X(\xi)>\lambda\text{ or }Y(\xi)>\lambda\right]\leq\frac2\lambda<1.
\]
Choose $\xi_0$ outside this event.
Then $\E_G\|f(G;\xi_0)\|^2\leq \lambda\gamma k$, and its failure probability is at most $\lambda\delta$; if $\lambda\delta>1$, the trivial bound $q_{\mathrm{fail}}(\xi_0)\leq1$ gives the stated minimum.

If $\delta=0$, nonnegativity and $\E_\xi q_{\mathrm{fail}}=0$ imply that $q_{\mathrm{fail}}(\xi)=0$ for almost every seed, whereas Markov's inequality gives $\Prb_\xi[X(\xi)\leq \lambda]\geq1-1/\lambda>0$.
These two sets intersect, so we may choose $\xi_0$ in their intersection.
In either case, fixing the seed preserves the degree bound and the tolerance $\eta$.
\end{proof}

\subsection{Algorithmic hardness of low-degree polynomials}

We are now ready to prove Theorem~\ref{thm:sparse_low_degree_hardness}.
We first dispose of the case $\eps>1/2$.
In this case, take any fixed $c_1,c_2>0$ and put
\[
    \alpha=\frac{\eps-1/2}{2}>0.
\]
Then $(1+\alpha)\Phi<(1/2+\eps)\Phi$, so Lemma~\ref{lem:sparse_largest_IM}, applied with $\alpha$ in place of $\eps$, shows that $G(n,d/n)$ has an induced matching of the target size with probability at most $e^{-cn}$.
On the other hand, an optimizer in the theorem's range would succeed with probability at least $1-\delta\geq1-n^{-c_2}$, because $\gamma D\geq1$.
These bounds are incompatible for large $n$.

Assume henceforth that $0<\eps\leq1/2$, and suppose, for contradiction, that a random polynomial as in the theorem exists, where the constants $c_1,c_2$ are those given by Lemma~\ref{lem:hardness_good_events} below.
Choose an integer $K\geq 4/\eps^2$, and set
\[
    T=(K-1)m,
    \qquad\text{and}\qquad
    \theta=\frac{\eps}{192\gamma(1+\eps)}.
\]
For fixed $K$, we have $T\leq n^3$ for all sufficiently large $n$, so we apply Proposition~\ref{prop:sparse_endpoint_forbidden_structure} with exponent $A=3$, and take $d$ large enough for this proposition and for Lemmas~\ref{lem:sparse_largest_IM} and~\ref{lem:sparse_endpoint_avoidance_checkpoint}.
Since $\delta\leq n^{-c_2\gamma D}$ and $\gamma D\geq1$, we have $3\delta<1$ for all sufficiently large $n$.
Lemma~\ref{lem:random_poly_to_deterministic}, with $\lambda=3$, therefore gives a deterministic polynomial $f$ of degree at most $D$ satisfying
\begin{equation}\label{eq:sparse_hardnessProof_deterministic_bounds}
    \E_{G} \norm{f(G)}^2 \leq 3\gamma\inparen{\frac{1}{2}+\eps}\Phi
    \qquad\text{and}\qquad
    \Prb_G\insquare{\abs{\mathcal M_f^\eta(G)} < \inparen{\frac{1}{2} + \eps}\Phi } \leq 3\delta.
\end{equation}
We have $\mathcal E_f>0$: since the failure probability in~\eqref{eq:sparse_hardnessProof_deterministic_bounds} is less than one, the rounded output is nonempty with positive probability, and a nonempty output contains an edge $e$ with $f_e(G)\geq1$.
Also, $d/n\leq1/2$ for all sufficiently large $n$, since $d$ is fixed.
Thus Proposition~\ref{prop:sparse_low_degree_stability} applies.

Henceforth fix this $f$.
Let $G^{(0)},\dots,G^{(T)}$ be an interpolation path as in Definition~\ref{def:sparse_interpolation_path}, and let
\[
    U_t\coloneqq\mathcal M_f^\eta(G^{(t)}).
\]

We first describe the contradiction at a high level.
The goal is to build sets $S_1,\dots,S_K$ from the outputs $U_t$ and show that they form the forbidden structure of Proposition~\ref{prop:sparse_endpoint_forbidden_structure}.
Proposition~\ref{prop:sparse_low_degree_stability} provides an event of non-negligible probability on which the output of $f$ changes little when any single edge is resampled.
On this event, the induced matchings $U_t$ change only gradually along the interpolation path; in particular, their endpoint sets cannot acquire many new vertices in a single step.
On the other hand, Lemma~\ref{lem:sparse_endpoint_avoidance_checkpoint} shows that, after all edges have been refreshed, a large induced matching is unlikely to be supported almost entirely on a fixed set of old endpoints of size $O(\Phi)$.
Thus the outputs along the path are eventually forced to introduce many new endpoints.
Combining these two facts, we extract matchings $S_1,\dots,S_K$ satisfying properties \ref{item:forbidden-time}--\ref{item:forbidden-new} of Proposition~\ref{prop:sparse_endpoint_forbidden_structure}, which yields the desired contradiction.
We first set up the selection rule and the relevant events, and then carry out this plan in Lemmas~\ref{lem:hardness_good_events}--\ref{lem:hardness_selection}.

Let
\[
    \cF_t\coloneqq\sigma(G^{(0)},Z_1,\ldots,Z_t).
\]
If $\tau$ is an almost surely finite integer-valued stopping time for $(\cF_t)$, then $G^{(\tau+m)}$ is independent of $\cF_\tau$ and has law $G(n,d/n)$, because the next $m$ steps replace every coordinate once by fresh variables.

Following the proof of~\cite[Theorem~1.3]{Wein2021}, set $t_1=0$ and $S_1=U_0$.
Given $S_1,\dots,S_{j-1}$ and $t_{j-1}$ for some $2\leq j\leq K$, let
\[
    W_{<j}\coloneqq\bigcup_{\ell<j}V(S_\ell),
\]
and define
\[
    t_j\coloneqq\min\left(
    \left\{t\in[t_{j-1},t_{j-1}+m]:
    \abs{V(U_t)\setminus W_{<j}}\geq\frac{\eps}{4}\Phi
    \right\}
    \cup\{t_{j-1}+m\}\right)
    \qquad\text{and}\qquad
    S_j\coloneqq U_{t_j}.
\]
In particular, $t_j=t_{j-1}+m$ if the threshold is not reached.
Since $W_{<j}$ is $\cF_{t_{j-1}}$-measurable and $(U_t)$ is adapted, $t_j$ is a stopping time, and inductively $t_j\leq(j-1)m$.
Put $B\coloneqq2K(1+\eps)$, and define
\[
    \widehat W_{<j}\coloneqq
    \begin{cases}
        W_{<j},&|W_{<j}|\leq B\Phi,\\
        \varnothing,&|W_{<j}|>B\Phi.
    \end{cases}
\]
This set is $\cF_{t_{j-1}}$-measurable and always has at most $B\Phi$ vertices.

We consider the following four events:
\begin{enumerate}[label=$\mathsf E_{\arabic*}$:, ref=$\mathsf E_{\arabic*}$]
    \item\label{item:hardness-size} $\inparen{\frac12+\eps}\Phi\leq\abs{U_t}\leq(1+\eps)\Phi$ for every $0\leq t\leq T$;
    \item\label{item:hardness-stable} no interpolation step is $\theta$-bad for $f$ (see Definition~\ref{def:sparse_bad_update});
    \item\label{item:hardness-refresh} for every $2\leq j\leq K$, every induced matching $M$ in $G^{(t_{j-1}+m)}$ with $\inparen{\frac12+\eps}\Phi\leq \abs{M}\leq (1+\eps)\Phi$ satisfies
    \begin{equation}\label{eq:sparse_hardnessProof_endpoint_refresh}
        \abs{V(M)\setminus \widehat W_{<j}}>\frac{\eps}{2}\Phi;
    \end{equation}
    \item\label{item:hardness-C} the conclusion of Proposition~\ref{prop:sparse_endpoint_forbidden_structure} holds, that is, no sets $S_1',\dots,S_K'$ along this interpolation path satisfy properties \ref{item:forbidden-time}--\ref{item:forbidden-new} of the proposition.
\end{enumerate}
We begin by arguing that all the above events occur simultaneously with positive probability.

\begin{lemma}\label{lem:hardness_good_events}
    There exist constants $c_1,c_2>0$, depending only on $\eps$, $d$, and $K$, such that the following holds.
    If $\delta\leq n^{-c_2\gamma D}$ and $\gamma D\log n\leq c_1 n$, then, for all sufficiently large $n$,
    \[
        \Prb\insquare{\mathsf E_1\cap\mathsf E_2\cap\mathsf E_3\cap\mathsf E_4}>0.
    \]
\end{lemma}

\begin{proof}
By Proposition~\ref{prop:sparse_low_degree_stability} with $\ell=K-1$, there is a constant $a_0=a_0(\eps,K)>0$ such that
\[
    \Prb\insquare{\mathsf E_2}
    \geq
    \inparen{\frac dn}^{4(K-1)D/\theta}
    \geq
    \exp(-a_0\gamma D\log n).
\]
By~\eqref{eq:sparse_hardnessProof_deterministic_bounds}, Lemma~\ref{lem:sparse_largest_IM}, and a union bound over $0\leq t\leq T$,
\begin{equation}\label{eq:sparse_hardnessProof_size_event}
    \Prb\insquare{\mathsf E_1^c}
    \leq
    3\delta(T+1)+\exp(-\Omega(n)).
\end{equation}
For each $2\leq j\leq K$, the graph $G^{(t_{j-1}+m)}$ is independent of $\cF_{t_{j-1}}$ and has law $G(n,d/n)$, while $\widehat W_{<j}$ is $\cF_{t_{j-1}}$-measurable with at most $B\Phi$ vertices.
We may therefore apply Lemma~\ref{lem:sparse_endpoint_avoidance_checkpoint} conditionally on $\cF_{t_{j-1}}$, with $W=\widehat W_{<j}$ and $\mu=\eps/2<2(1/2+\eps)^2$.
A union bound over $j=2,\dots,K$ gives $\Prb\insquare{\mathsf E_3^c}\leq e^{-\Omega(n)}$.
Finally, Proposition~\ref{prop:sparse_endpoint_forbidden_structure} gives $\Prb\insquare{\mathsf E_4^c}\leq e^{-\Omega(n)}$.

Combining these bounds, there is a fixed $c=c(\eps,d,K)>0$ such that
\[
    \Prb\insquare{\mathsf E_1^c\cup\mathsf E_3^c\cup\mathsf E_4^c}
    \leq
    3\delta(T+1)+e^{-cn}.
\]
For large $n$, we have $3(T+1)\leq n^4$.
Choose $c_2>a_0+6$.
Since $\gamma D\geq1$,
\[
    3\delta(T+1)
    \leq
    n^4e^{-c_2\gamma D\log n}
    \leq\frac14e^{-a_0\gamma D\log n}
\]
for all sufficiently large $n$.
Moreover, choosing $c_1>0$ so small that $a_0c_1<c/2$, the assumption $\gamma D\log n\leq c_1n$ gives
\[
    e^{-cn}\leq\frac14e^{-a_0\gamma D\log n}
\]
for all sufficiently large $n$.
Consequently, $\Prb\insquare{\mathsf E_2}$ exceeds $\Prb\insquare{\mathsf E_1^c\cup\mathsf E_3^c\cup\mathsf E_4^c}$, and the lemma follows.
\end{proof}

Next, we show that $\abs{U_t\triangle U_{t-1}}$ is small for each $t$ under the event $\mathsf E_1\cap\mathsf E_2$.

\begin{lemma}\label{lem:hardness_small_steps}
    On the event $\mathsf E_1\cap\mathsf E_2$, for every $1\leq t\leq T$,
    \begin{equation}\label{eq:sparse_hardnessProof_Ut_UtminusOne_symmDiff}
        \abs{U_t\triangle U_{t-1}}
        \leq
        \frac{\eps}{8}\Phi
    \end{equation}
    for all sufficiently large $n$.
\end{lemma}

\begin{proof}
By~\ref{item:hardness-size}, $\abs{U_t}\geq(\frac12+\eps)\Phi$, so the failure case in Definition~\ref{def:rounded-polynomial-output} does not occur at any time $t$.
Fix $1\leq t\leq T$, and for $s\in\{t-1,t\}$, write $A_s,\tilde A_s,B_s$ for the sets in Definition~\ref{def:rounded-polynomial-output} evaluated at $G^{(s)}$.
Apart from possibly the coordinate resampled at time $t$, each $e\in U_t\triangle U_{t-1}$ falls into one of the following two cases:
\begin{enumerate}[(a)]
    \item one of $f_e(G^{(t)})$ and $f_e(G^{(t-1)})$ is at least $1$ and the other is at most $1/2$;
    \item $e\in(A_s\setminus\tilde A_s)\cup B_s$ for some $s\in\{t-1,t\}$.
\end{enumerate}
In particular, an edge that remains a candidate but is removed because of a conflict is counted in~(b).
Since the failure case does not occur, $\abs{(A_s\setminus\tilde A_s)\cup B_s}\leq\eta m$ for each $s\in\{t-1,t\}$.
Hence at least
\[
    \abs{U_t\triangle U_{t-1}}-2\eta m-1
\]
coordinates fall into case (a), and each of them contributes at least $1/4$ to $\norm{f(G^{(t)})-f(G^{(t-1)})}^2$.
Therefore,
\[
    \norm{f(G^{(t)})-f(G^{(t-1)})}^2
    \geq
    \frac14\inparen{\abs{U_t\triangle U_{t-1}}-2\eta m-1}.
\]
On the other hand, by~\ref{item:hardness-stable}, Definition~\ref{def:sparse_bad_update}, and~\eqref{eq:sparse_hardnessProof_deterministic_bounds},
\[
    \norm{f(G^{(t)})-f(G^{(t-1)})}^2
    \leq
    \theta\,\E_G\norm{f(G)}^2
    \leq
    3\theta\gamma\inparen{\frac12+\eps}\Phi.
\]
Combining these inequalities gives
\[
    \abs{U_t\triangle U_{t-1}}
    \leq
    12\theta\gamma\inparen{\frac12+\eps}\Phi+2\eta m+1
    \leq
    \frac{\eps}{16}\Phi+\frac{\eps}{32}\Phi+1
    \leq
    \frac{\eps}{8}\Phi,
\]
where the middle inequality uses the choices of $\theta$ and $\eta$, and the last holds for all sufficiently large $n$, since $\Phi=\Theta_d(n)$.
\end{proof}

Next, we argue that, on the event  $\mathsf E_1\cap\mathsf E_2\cap\mathsf E_3$, we can construct sets $S_1,\dots,S_K$ satisfying properties \ref{item:forbidden-time}--\ref{item:forbidden-new}, contradicting Proposition~\ref{prop:sparse_endpoint_forbidden_structure}.

\begin{lemma}\label{lem:hardness_selection}
    On the event $\mathsf E_1\cap\mathsf E_2\cap\mathsf E_3$, for all sufficiently large $n$, the sets $S_1,\dots,S_K$ produced by the selection rule satisfy properties \ref{item:forbidden-time}--\ref{item:forbidden-new} of Proposition~\ref{prop:sparse_endpoint_forbidden_structure}.
\end{lemma}

\begin{proof}
We show by induction on $j$ that $t_j\leq(j-1)m$, and that, for $j\geq2$,
\begin{equation}\label{eq:sparse_hardnessProof_new_endpoints}
    \abs{V(S_j)\setminus W_{<j}}
    \in
    \insquare{\frac{\eps}{4},\frac{\eps}{2}}\Phi.
\end{equation}
The case $j=1$ is immediate, since $t_1=0$.
Suppose $2\leq j\leq K$ and that the previous steps have produced $S_1,\dots,S_{j-1}$ and $t_{j-1}\leq (j-2)m$.
By~\ref{item:hardness-size},
\[
    |W_{<j}|\leq2\sum_{\ell<j}|S_\ell|
    \leq2K(1+\eps)\Phi=B\Phi,
\]
so $\widehat W_{<j}=W_{<j}$.
Since $t_{j-1}+m\leq (j-1)m\leq T$, the induced matching $U_{t_{j-1}+m}$ has between $\inparen{\frac12+\eps}\Phi$ and $(1+\eps)\Phi$ edges by~\ref{item:hardness-size}.
By~\ref{item:hardness-refresh}, applied at step $j$ with $M=U_{t_{j-1}+m}$,
\[
    \abs{V(U_{t_{j-1}+m})\setminus W_{<j}}
    >
    \frac{\eps}{2}\Phi.
\]
On the other hand, $V(U_{t_{j-1}})=V(S_{j-1})\subseteq W_{<j}$, so no endpoint of $U_{t_{j-1}}$ lies outside $W_{<j}$.
Thus the threshold $\eps\Phi/4$ is reached in $[t_{j-1},t_{j-1}+m]$, and $t_j>t_{j-1}$.
By the definition of $t_j$,
\[
    \abs{V(U_{t_j-1})\setminus W_{<j}}
    <\frac{\eps}{4}\Phi,
    \qquad\text{and}\qquad
    \abs{V(U_{t_j})\setminus W_{<j}}
    \geq\frac{\eps}{4}\Phi.
\]
Moreover, $t_j\leq t_{j-1}+m\leq (j-1)m$ and $S_j=U_{t_j}$.
Using the strict inequality above together with Lemma~\ref{lem:hardness_small_steps},
\[
    \abs{V(S_j)\setminus W_{<j}}
    \leq
    \frac{\eps}{4}\Phi
    +
    \abs{V(U_{t_j})\setminus V(U_{t_j-1})}
    \leq
    \frac{\eps}{4}\Phi
    +
    2\abs{U_{t_j}\triangle U_{t_j-1}}
    \leq
    \frac{\eps}{2}\Phi,
\]
which proves~\eqref{eq:sparse_hardnessProof_new_endpoints} and completes the induction.

Since $W_{<j}=\bigcup_{\ell<j}V(S_\ell)$, the bound~\eqref{eq:sparse_hardnessProof_new_endpoints} is property~\ref{item:forbidden-new}.
Moreover, by~\ref{item:hardness-size}, each $S_j=U_{t_j}$ is an induced matching in $G^{(t_j)}$ with at least $(\frac12+\eps)\Phi$ edges, where $t_j\leq (K-1)m=T$, which gives properties~\ref{item:forbidden-time} and~\ref{item:forbidden-size}.
\end{proof}

We can now complete the proof of Theorem~\ref{thm:sparse_low_degree_hardness}.
By Lemma~\ref{lem:hardness_good_events}, for all sufficiently large $n$, the events $\mathsf E_1,\dots,\mathsf E_4$ occur simultaneously with positive probability.
On this intersection, Lemma~\ref{lem:hardness_selection} shows that $S_1,\dots,S_K$ satisfy properties \ref{item:forbidden-time}--\ref{item:forbidden-new} of Proposition~\ref{prop:sparse_endpoint_forbidden_structure}, contradicting~\ref{item:hardness-C}.
This completes the proof.

\section*{AI methodology statement}
The authors formulated the problem, identified the factor-two gap, and developed the setup of the online and low-degree polynomial algorithms. They then proved lower bounds for online algorithms and low-degree polynomials, along with an achievability result for online algorithms in a range of densities $p$. 

The authors then started querying OpenAI's ChatGPT-5.5 initially to extend the online algorithmic guarantee to a broader range of densities $p$. While constructing the proofs for the achievability and the barrier for online algorithms following existing techniques \cite{GamarnikKizildagWarnke2025}, the authors used LLMs also to explore simpler analyses of online algorithms through an interactive process in Codex. This resulted in the proof of the online achievability through maximality and the direct proof of the online barrier without using the overlap gap property, both of which are simpler than the existing techniques \cite{GamarnikKizildagWarnke2025, dhawan2025sharp, dhawan2026algorithmic} as well as the original proof ideas the authors came up with.

As for the low-degree polynomial algorithms, the authors conceived the proof idea of finding independent sets in the square of the line graph of $G$ using constant-degree polynomials. 
Under the authors' guidance, ChatGPT-5.6 then formalized this idea via the edge coloring approach. The authors then developed the approach to obtain the eventual proof.

The authors redeveloped all AI-generated content and wrote the entire manuscript. They take full responsibility for the final content.

\bibliographystyle{alpha}
\bibliography{ref}

\end{document}